\documentclass[fleqn,usenatbib]{mnras}

\usepackage{newtxtext,newtxmath}

\usepackage[T1]{fontenc}

\DeclareRobustCommand{\VAN}[3]{#2}
\let\VANthebibliography\thebibliography
\def\thebibliography{\DeclareRobustCommand{\VAN}[3]{##3}\VANthebibliography}

\usepackage{graphicx}	
\usepackage{amsmath}	

\title[Gas accretion in Auriga Milky Way-like galaxies]{Cosmological gas accretion history onto the stellar discs of Milky Way-like galaxies in the Auriga simulations -- (I) Temporal dependency}

\author[F.~G. Iza et al.]{Federico~G.~Iza$^{1,2}$\thanks{E-mail: fiza@iafe.uba.ar},
Cecilia~Scannapieco$^{2}$,
Sebasti\'an~E.~Nuza$^{1,2}$,
Robert~J.~J.~Grand$^{3,4}$, \newauthor
Facundo~A.~Gómez$^{5,6}$,
Volker~Springel$^{7}$,
Rüdiger~Pakmor$^{7}$,
Federico Marinacci$^{8}$
\\
\\
$^1$ Instituto de Astronom\'ia y F\'isica del Espacio (IAFE, CONICET-UBA), 1428 Buenos Aires, Argentina \\
$^2$ Departamento de Física, FCEyN, Universidad de Buenos Aires, CONICET, Ciudad Universitaria, 1428 Buenos Aires, Argentina \\
$^3$ Instituto de Astrof\'isica de Canarias, Calle V\'ia L\'actea s/n, E-38205 La Laguna, Tenerife, Spain \\
$^4$ Departamento de Astrofísica, Universidad de La Laguna, Av. del Astrofísico Francisco Sánchez s/n, E-38206 La Laguna, Tenerife, Spain \\
$^5$ Departamento de Astronom\'ia, Universidad de La Serena, Avenida Juan Cisternas 1200, La Serena, Chile \\
$^6$ Instituto de Investigaci\'on Multidisciplinar en Ciencia y Tecnolog\'ia, Universidad de La Serena, Ra\'ul Bitr\'an 1305, La Serena, Chile \\
$^7$ Max-Planck-Institut für Astrophysik, Karl-Schwarzschild-Str 1, D-85748 Garching, Germany \\
$^8$ Department of Physics \& Astronomy, University of Bologna, via Gobetti 93/2, I-40129 Bologna, Italy
}

\date{Accepted XXX. Received YYY; in original form ZZZ}

\pubyear{2022}

\begin{document}
\label{firstpage}
\pagerange{\pageref{firstpage}--\pageref{lastpage}}
\maketitle

\begin{abstract}
We use the 30 simulations of the Auriga Project to estimate the temporal dependency of the inflow, outflow and net accretion rates onto the discs of Milky Way-like galaxies.
The net accretion rates are found to be similar for all galaxies at early times, increasing rapidly up to $\sim 10~\mathrm{M}_\odot \, \mathrm{yr}^{-1}$.
After $\sim 6~\mathrm{Gyr}$ of evolution, however, the net accretion rates are diverse: in most galaxies, these exhibit an exponential-like decay, but some systems instead present increasing or approximately constant levels up to the present time.
An exponential fit to the net accretion rates averaged over the MW analogues yields typical decay time-scale of $7.2~\mathrm{Gyr}$.
The analysis of the time-evolution of the inflow and outflow rates, and their relation to the star formation rate (SFR) in the discs, confirms the close connection between these quantities.
First, the inflow$/$outflow ratio stays approximately constant, with typical values of $\dot{M}_\mathrm{out}/ \dot{M}_\mathrm{in} \sim 0.75$, indicating that the gas mass involved in outflows is of the order of 25\% lower compared to that involved in inflows.
A similar behaviour is found for the SFR$/$inflow rate ratio, with typical values between 0.1 and 0.3, and for the outflow rate$/$SFR which varies in the range $3.5$--$5.5$.
Our results show that continuous inflow is key to the SFR levels in disc galaxies, and that the star formation activity and the subsequent feedback in the discs is able to produce mass-loaded galaxy winds in the disc-halo interface.
\end{abstract}




\section{Introduction}

A key process in the formation and evolution of galaxies in the context of the Lambda Cold Dark Matter ($\Lambda$CDM) model is the accretion of gas, an ubiquitous and continuous mechanism through which dark matter haloes obtain fresh gaseous material that may eventually serve as fuel for star formation \citep[e.g.][]{Putman2012, Nuza2014, Richter2017}.
Observations and theoretical studies indicate that such accretion can occur in various forms, including diffuse and filamentary inflow from the intergalactic medium \citep[][]{Keres2005, Brooks2009}, and from gas reservoirs linked to remnants of galaxy interactions \citep[e.g.][]{Richter2017, Zhu2021, Xu2021}.
On the other hand, accretion of gas onto the disc region can come directly from the circumgalactic medium (CGM), as a result of condensation of gaseous material in the outer halo \citep[][]{White1978} and/or in the form of galactic fountains \citep{Shapiro1976}.
The circulation of gas in the disc-halo interface (DHI) is therefore extremely complex, combining inflows from different channels and outflows generated by stellar winds, supernova explosions and active galactic nuclei.
In this context, the accretion, star formation and outflow rates are inter-related in a complex way during the formation and evolution of galactic systems \citep{Putman2012, Tumlinson2017}.

The sustained star formation activity over many gigayears of evolution seen in disc galaxies suggests that the exchange of gas resulting from inflowing and outflowing material should be such that inflows dominate over outflows.
In this way, galaxies would gain gas mass over time and replenish the star-forming material in the disc region.
During recent decades, observational evidence of the presence of extraplanar gas and ongoing accretion onto the central regions of nearby galaxies and the Milky Way (MW) has been found \citep[e.g.][]{Shull2009, Lehner2012, Richter2017, Bish2019, Marasco2019}.
In the MW, estimates of the present-day gas accretion rate were traditionally derived from inflowing gas clouds detected through the $21~\mathrm{cm}$ line in emission with velocities that are inconsistent with the rotation of the Galaxy, the so-called high-velocity clouds (HVCs).
After considering the contribution of gas complexes surrounding the Galaxy but excluding the Magellanic Stream (MS), typical accretion rates obtained from neutral HVCs lie in the range $\sim 0.1$--$0.4 ~\mathrm{M}_\odot \, \mathrm{yr}^{-1}$ \citep[][]{Putman2012}.
Similarly, by modelling the three-dimensional distribution of {\it all} known neutral HVCs (i.e. including the MS), \cite{Richter2012} estimated an accretion rate of $0.7 ~\mathrm{M}_\odot \, \mathrm{yr}^{-1}$ for MW/Andromeda-type galaxies.
Moreover, \cite{Lehner2012} have shown that the traditional HVCs tend to have ionized gas envelopes extending far from the observed H\,\textsc{i} contours harbouring at least as much mass as their neutral cores.
In this respect, \cite{Richter2017} derived an accretion rate of neutral and ionized material in the CGM traced by HVCs of $\sim 6 \, \mathrm{M}_\odot \, \mathrm{yr}^{-1}$, where the MS contributes with about 90\% of the mass inflow.
This material may feed the MW disc at values of a few solar masses per year over the next Gyr, consistent with the requirements of the Galaxy's current star formation rate (SFR).

More recently, the analysis of the properties of HVCs with large covering factors in the MW made it possible to estimate not only the inflow rate onto the central regions of the Galactic halo, but also the outflow rate separately.
The estimations of \cite{Fox2019} for the inflow and outflow rates of cold ionized gas at distances less than $\sim 12 ~\mathrm{kpc}$ from the Sun (therefore excluding the MS) are $\dot{M}_{\mathrm{in}} \gtrsim (0.53 \pm 0.31) ~\mathrm{M}_\odot \, \mathrm{yr}^{-1}$ and $\dot{M}_{\mathrm{out}} \gtrsim (0.16 \pm 0.10) ~\mathrm{M}_\odot \, \mathrm{yr}^{-1}$, after averaging over cloud metallicities.
These results suggest that the MW is currently in an inflow-dominated phase, although outflows are important as well.
As mentioned above, this {\it net} inflow of gas is necessary to feed the disc.
However, uncertainties in the role of the various inter-related processes affecting the DHI make it difficult to determine the amount of star-forming material that is actually able to reach the disc region.
According to \cite{RichterASSL}, if one assumes that half of the diffuse ionized gas believed to be present in the DHI is actually accreting, observational constraints for low-velocity material located at a few kpc above the disc imply rates of about 1--$2 ~\mathrm{M}_\odot \, \mathrm{yr}^{-1}$, thus in line with the required SFR.

From a theoretical point of view, numerous studies with cosmological, hydrodynamical simulations have investigated how gas is accreted onto haloes and galaxies, the properties of such gas and the effects of feedback and galactic outflows on the characteristics of the accretion (e.g. \citealt{Murali2002, Keres2005, Ocvirk2008, Nelson2013, Nuza2014, Christensen2016}).
These works found that gas accretion onto the central regions of galaxies occurs not only via the cooling of gas from the hot gaseous halo but also through filamentary accretion directly tunneled into the disc, where the relative importance of these `hot' and `cold' modes depends sensitively on halo mass and redshift (e.g. \citealt{Keres2005, Ocvirk2008, Dekel2009, Keres2009}).
The exact levels of inflow, outflow and net accretion onto galaxy discs are, however, not fully predicted by cosmological simulations, as the mixing of outflowing material resulting from various feedback channels and inflowing material from filaments and galactic fountains might be quite sensitive to the feedback prescription adopted and, to some extent, could also be affected by numerical issues (see, e.g., \citealt{Oppenheimer2010, Faucher2011, Scannapieco2012, Shen2012, Nelson2013}).

While the aforementioned simulations showed that the gas accretion levels depend on halo mass and redshift, environment might also play a role on the determination of the accretion rates onto the central region of galaxies, particularly in the case of filamentary accretion from the intergalactic medium.
This is important for studies of the MW which inhabits, together with its neighbour Andromeda, an overdense region of the Universe.
In this respect, \cite{Nuza2014} studied the distribution of gas in a MW-like environment using simulations of the Local Group (LG) performed within the context of the \textsc{clues}\footnote{\url{www.clues-project.org}} (Constrained Local UniversE Simulations) project.
In these simulations, a significant excess of gas was detected in the intergalactic region between the MW and Andromeda, compared to any other direction (see also \citealt{Damle2022} for an analysis of the gas distribution in a new generation of LG simulations).
Such gas excess, which seems to be consistent with observational findings towards the Andromeda's direction \citep[e.g.][]{Richter2017}, provides a possible indication that the accretion of gas might be different for galaxies in different environments.
Furthermore, using a set of similar simulations, \cite{Creasey2015} found that galaxies in LG-like regions have systematically higher SFRs than galaxies in less dense environments.
Owing to the central role of our MW as a benchmark for understanding the galaxy formation process, it is important that the accretion rates onto the disc region are investigated in more detail, as well as the relation between accretion and star formation and the effects of feedback on the circulation of gas in and around the disc.

Quantifying the amount of gas entering the disc region of spiral galaxies as a function of time and radius is also of prime importance for chemical evolution models (CEMs).
Such models intend to reconstruct the formation of the various stellar components -- thin/thick discs, bulge and stellar halo -- using a set of inputs constrained by observational information.
In the case of the disc, the most important assumptions relate to the SFR and the gas accretion law -- an analytic prescription that depends both on cosmic time and galactocentric distance accounting for the gas mass that needs to be added to the model galaxy owing to the lack of a cosmological context in CEMs -- and the constraints are various observations on the present-day chemical properties across the disc.
Despite its importance, a reliable accretion law is still not determined, but current models are able to reproduce the observational results if the accretion rate of gas is assumed to depend on time -- usually in an exponential-decay manner -- and on radius -- following an inside-out-like behaviour \citep[e.g.][]{Chiappini2001}.
For a recent review on the chemical evolution modelling of the MW see, for instance, \cite{Matteucci2021}.

Our main goal is to study the circulation of gaseous material onto the (stellar) disc region of galaxies in cosmological simulations.
Additionally, this information can be used to provide CEMs with a physically-motivated accretion law that is consistent with galaxy formation within the context of the $\Lambda$CDM model, as proposed by \cite{Nuza2019} using a sample of 4 simulated MW-like galaxies.
In this work, we use the simulations of the Auriga Project \citep{Grand2017}, a set of 30 MW-mass galaxies simulated with the magnetohydrodynamical, moving/mesh code \textsc{arepo} \citep{Springel2010}.
We also use 9 additional simulations that were run including tracer particles \citep{Genel2013}, which allows following the trajectories of the gas elements through cosmic time.
The relatively high number of simulated galaxies provides an ideal set for quantifying accretion levels onto the disc regions of simulated galaxies, as well as the expected galaxy-to-galaxy variations.
It is important to note that the simulated galaxies are consistent with many observed properties of the MW and similar spiral galaxies; however, the systems were selected to be relatively isolated at $z=0$ and, therefore, inhabit a different environment than the MW.
In the present paper we focus our analysis on the temporal evolution of the inflow, outflow and net accretion rates, while separate works follow up with the investigation of the corresponding radial dependencies and of possible environmental effects affecting galaxies in LG-like environments. 

This paper is organized as follows: in Section~\ref{sec:simulations} we present the sample of 30 simulated galaxies used in this work, including the sub-sample of 9 resimulations using tracer particles; in Section~\ref{sec:analysis} we define the disc region and introduce the method used to compute incoming/outcoming gas flows of the galaxy discs; in Section~\ref{sec:results} we present our results for the inflow, outflow and net accretion rates and analyse their relation with the SFR; and in Section~\ref{sec:MWanalogs} we split the sample in two groups to focus on the accretion law of our MW-like analogues.
Finally, in Section~\ref{sec:conclusions}, we present a discussion and our conclusions.

\section{Simulations}
\label{sec:simulations}

\subsection{The Auriga galaxy sample}

In this work, we analyse 30 galaxies from the Auriga Project \citep{Grand2017}, a set of high-resolution, zoom-in cosmological simulations performed with the magnetohydrodynamic (MHD) code \textsc{arepo} \citep{Springel2010}.
The latter is a quasi-Lagrangian, moving-mesh code that follows the evolution of MHD and collisionless dynamics in a cosmological environment.
Gravitational forces are computed using a standard TreePM treatment and MHD equations are calculated with a second-order Runge-Kutta method on a dynamic mesh constructed from a Voronoi tessellation.

The galaxy formation model used in the Auriga Project includes primordial and metal-line cooling, a uniform ultraviolet (UV) background field for reionization, star formation (as in \citealt{Springel2003}, with a density threshold of $0.13~\mathrm{cm}^{-3}$ and a star formation time-scale of $\tau=2.2~\mathrm{Gyr}$), magnetic fields \citep{Pakmor2014, Pakmor2017, Pakmor2018}, active galactic nuclei, energetic and chemical feedback from Type II supernovae, and mass loss/metal return owing to Type Ia supernovae and asymptotic giant branch stars \citep{Vogelsberger2013, Marinacci2014, Grand2017}.

Star particles in the simulation characterise a single stellar population (SSP) with a given age and metallicity.
For a SSP, the number of Type Ia supernovae events is calculated by integrating the delay time distribution function (DTD) as indicated in \cite{Grand2017}.
The amount of mass and metals injected into the ISM is then calculated from SNIa yield tables \citep{Thielemann2003, Travaglio2004} and distributed among neighbouring gas cells.

Type II supernova events are assumed to occur instantaneously and are implemented by transforming a probabilistically chosen star-forming gas cell into either a star or a wind particle.
These particles are loaded with 0.4 times the metal mass of the gas cells from which they are created and launched applying a velocity kick proportional to the local 1D dark matter velocity dispersion. The particles receiving kicks travel until they reach a cell with a density below 0.05 times the physical density threshold for star formation, or if a maximum travel time of 0.025 times the Hubble time at the current time-step is reached.
Then, the wind particles are dissolved and deposit their mass, momentum, thermal energy and metals into the gas cell in which they are located.

\begin{table*}
	\centering
	\caption{Galactic properties at $z=0$. The columns are: (1) galaxy name, (2) virial radius $R_{200}$, (3) virial mass $M_{200}$, (4) subhalo stellar mass $M_\star$, (5) subhalo gas mass $M_\mathrm{gas}$, (6) subhalo baryon mass $M_\mathrm{b}$, (7) disc-to-total mass fraction, (8) disc radius $R_\mathrm{d}$, (9) disc height $h_\mathrm{d}$, and (10) group of each galaxy as defined in Section~\ref{sec:MWanalogs} (G1: group 1, G2: group 2, E: excluded). The symbol $\dagger$ in the first column identifies galaxies that have been re-simulated including a treatment for stochastic tracer particles. The methods used to calculate quantities in columns (7), (8) and (9) are described in Section~\ref{sec:analysis}.}
	\label{tab:galactic_properties}
	\begin{tabular}{lcccccccccc}
		\hline
		Galaxy 	& $R_{200}$ [kpc] & $M_{200}$ [$10^{10}~\mathrm{M}_\odot$] & $M_\star$ [$10^{10}~\mathrm{M}_\odot$] & $M_\mathrm{gas}$ [$10^{10}~\mathrm{M}_\odot$] & $M_\mathrm{b}$ [$10^{10}~\mathrm{M}_\odot$] & D/T 	& $R_\mathrm{d}$ [kpc] 	& $h_\mathrm{d}$ [kpc] & Group \\
		\hline
		Au1 	        & 206.0		& 93.4		& 3.0		& 8.8   & 11.8  & 0.76	& 19.7  & 2.7 & E \\
		Au2 	        & 261.7		& 191.4		& 9.4		& 12.5  & 21.9  & 0.81	& 33.7	& 3.1 & G1 \\
		Au3 	        & 239.0		& 145.8		& 8.7		& 9.7   & 18.5  & 0.74	& 23.6	& 2.5 & G1 \\
		Au4 	        & 236.3		& 140.9		& 8.8		& 12.8  & 21.6  & 0.38	& 21.8	& 3.6 & G2 \\
		Au5$^\dagger$   & 223.1		& 118.6		& 7.1		& 8.5   & 15.6  & 0.63	& 13.8	& 2.0 & E \\
		Au6$^\dagger$ 	& 213.8		& 104.4		& 5.4		& 6.8   & 12.2  & 0.80	& 21.1	& 2.5 & G1 \\
		Au7 	        & 218.9		& 112.0		& 5.9		& 11.6  & 17.5  & 0.60	& 21.7	& 3.2 & G2 \\
		Au8 	        & 216.3		& 108.1		& 4.0		& 9.5   & 13.4  & 0.84	& 29.2	& 3.5 & G1 \\
		Au9$^\dagger$ 	& 214.2		& 105.0		& 6.3		& 6.6   & 12.9  & 0.66	& 12.0	& 1.9 & G1 \\
		Au10 	        & 214.0		& 104.7		& 6.2		& 8.6   & 14.8  & 0.73	& 8.4	& 1.6 & G1 \\
		Au11 	        & 249.0		& 164.9		& 6.4		& 8.8   & 15.3  & 0.62	& 19.3	& 2.5 & G1 \\
		Au12 	        & 217.1		& 109.3		& 6.6		& 8.8   & 15.4  & 0.68	& 14.9	& 2.6 & G2 \\
		Au13$^\dagger$ 	& 223.3		& 118.9		& 6.7		& 10.0  & 16.7  & 0.68	& 12.3	& 2.4 & G1 \\
		Au14 	        & 249.4		& 165.7		& 11.7		& 14.1  & 25.8  & 0.61	& 17.7	& 2.8 & G1 \\
		Au15 	        & 225.4		& 122.2		& 4.3		& 9.3   & 13.7  & 0.69	& 19.2	& 3.0 & G2 \\
		Au16 	        & 241.4		& 150.3		& 7.0		& 10.4  & 17.4  & 0.88	& 31.1	& 3.1 & G1 \\
		Au17$^\dagger$ 	& 212.7		& 102.8		& 7.9		& 6.1   & 14.0  & 0.76	& 11.8	& 1.8 & G1 \\
		Au18 	        & 225.3		& 122.1		& 8.4		& 7.1   & 15.5  & 0.80	& 14.0	& 2.0 & G1 \\
		Au19 	        & 224.5		& 120.9		& 6.2		& 9.7   & 15.8  & 0.47	& 22.0	& 3.1 & E \\
		Au20 	        & 227.0		& 124.9		& 5.6		& 14.2  & 19.8  & 0.72	& 24.1	& 3.3 & G2 \\
		Au21 	        & 238.6		& 145.1		& 8.7		& 11.8  & 20.5  & 0.80	& 18.6	& 2.9 & G1 \\
		Au22 	        & 205.5		& 92.6		& 6.2		& 3.6   & 9.8   & 0.69	& 7.9	& 1.6 & G1 \\
		Au23$^\dagger$ 	& 245.2		& 157.5		& 9.8		& 9.8   & 19.6  & 0.77	& 18.0	& 2.2 & G1 \\
		Au24$^\dagger$ 	& 240.8		& 149.2		& 7.7		& 10.5  & 18.2  & 0.69	& 25.0	& 2.9 & G1 \\
		Au25 	        & 225.3		& 122.1		& 3.7		& 7.9   & 11.6  & 0.86	& 22.8	& 2.9 & G1 \\
		Au26$^\dagger$ 	& 244.6		& 156.4		& 11.4		& 12.9  & 24.3  & 0.71	& 10.7	& 1.8 & G1 \\
		Au27 	        & 253.8		& 174.5		& 10.3		& 12.5  & 22.8  & 0.71	& 17.2	& 2.4 & G1 \\
		Au28$^\dagger$ 	& 246.8		& 160.5		& 11.0		& 12.3  & 23.3  & 0.59	& 10.8	& 2.2 & E \\
		Au29 	        & 243.5		& 154.2		& 10.4		& 8.7   & 19.1  & 0.12	& 18.1	& 3.7 & E \\
		Au30 	        & 218.1		& 110.8		& 5.4		& 9.9   & 15.3  & 0.41	& 23.8	& 3.0 & E \\
		\hline
	\end{tabular}
\end{table*}

The mass resolution of the simulations is $\sim 3 \times 10^5\,\mathrm{M}_\odot$ and $\sim 5\times 10^4\,\mathrm{M}_\odot$ for dark matter and baryons, respectively, corresponding to the level 4 resolution runs of \cite{Grand2017}.
The softening length for star and dark matter particles is fixed in comoving coordinates at $500\,h^{-1}\,\mathrm{cpc}$ up to $z=1$; for later times the softening is constant and set to $369~\mathrm{pc}$ in physical coordinates.
The cosmological parameters assumed in the simulations are $\Omega_{\rm M} = 0.307$, $\Omega_{\rm b} = 0.048$, $\Omega_\Lambda = 0.693$ and a Hubble constant of $H_0 = 100\,h~\mathrm{km}\,\mathrm{s}^{-1}\,\mathrm{Mpc}^{-1}$ ($h = 0.6777$), in agreement with the cosmological parameter estimations of \cite{Planck2014}.
For each Auriga galaxy, there are a total of 128 snapshot files available, sampling their whole evolution (separated on average by $\sim 100~\mathrm{Myr}$).

The host haloes of the Auriga galaxies were chosen at $z=0$ from a parent dark matter-only cosmological simulation performed within a box of $100~\mathrm{cMpc}$ on a side \citep{Schaye2015}.
Two selection criteria were considered for the host haloes: to have virial masses in the range 1--$2 \times 10^{12} \, \mathrm{M}_\odot$, and to be relatively isolated (the center of the host must lie outside 9 times the virial radius of any other halo with a mass higher than 3\% of the host halo mass).
The 30 galaxies of the Auriga Project, labelled by the Auriga prefix ``Au'' followed by a number from 1 to 30, were randomly selected from the most isolated quartile.

The selected galaxies have $z=0$ virial masses that lie in the range $\sim 9$--$17~\times10^{11} \, \mathrm{M}_\odot$ and stellar masses of $\sim 3$--$12~\times10^{10} \, \mathrm{M}_\odot$, as shown in Table~\ref{tab:galactic_properties}.
The virial masses of the selected galaxies are similar to the commonly accepted value of $\sim 10^{12}~\mathrm{M}_\odot$ for the MW, and can be thought of as ``MW-like'' galaxies, although they have been chosen to be relatively isolated at $z=0$, in contrast to the real environment of the MW.
The table also provides information on the $z=0$ values of the virial radii ($R_{200}$), which lie in the range $\sim 206$--$262~\mathrm{kpc}$.

\subsection{Simulations using tracer particles}

As a result of the quasi-Lagrangian nature of \textsc{arepo}, it is not possible to follow the trajectories of gas elements through time, which would provide relevant information for our study on gas accretion.
However, a subset of the Auriga galaxies has been re-simulated including the so-called tracer particles, which enables us to follow the evolutionary history of gas cells \citep{Genel2013, DeFelippis2017}.
The tracers used in the Auriga galaxies are stochastic tracer particles, in contrast with tracers that follow the local fluid velocity; in practice, these tracers move with a probability that is given by the advection of mass through the boundary of each cell.
Tracer particles can then be followed in time, as in standard Lagrangian codes.

In particular, 9 of the Auriga galaxies have been re-simulated using tracer particles, saving 252 snapshots with an average time spacing of $\sim 55~\mathrm{Myr}$, although, for the purposes of this work, we keep the temporal spacing approximately constant at $\sim 100~\mathrm{Myr}$.
It is worth noting that for these 9 galaxies, both the original simulation and the one with tracer particles are available, which allows us to validate the results obtained with the former, in the case of the calculation of the net accretion rates.
More details about the simulations with tracer particles can be found in \cite{Grand2019}.

\section{Analysis}
\label{sec:analysis}

\subsection{Defining the disc region}
\label{sec:disc_radius_height}

The main objective of this work is to calculate the gas accretion rates, as a function of time\footnote{Throughout this work, we use the term "Time" to refer to cosmic time, not to be confused with "Lookback time".}, onto the discs of MW-mass galaxies.
This requires a proper identification of the disc region, as well as an evaluation of the morphological properties of the simulated galaxies, not only at $z=0$ but throughout their whole evolution.
With this purpose, we first calculate, for each galaxy and time, the inertia tensor of all the stars in the inner $10~\mathrm{ckpc}$ and rotate the reference system such that the principal axis lies in the $z$-direction, in a way that $z$ is also the direction of the angular momentum vector of the stars.
The galactic disc is therefore contained in the $xy$ plane.

To calculate the disc region of the simulated galaxies, we used a simple definition using two parameters: a disc radius ($R_{\rm d}$) and a disc height ($h_{\rm d}$).
The disc radius is defined, at each time, as the radius enclosing 90\% of the host stellar mass\footnote{Our choice of enclosing 90\% of the stellar mass is adequate for all galaxies, avoiding problems that might appear during merger events where the stellar distribution is perturbed.
We have checked that this method properly identifies the disc region, and that the exact value assumed for the enclosed mass does not affect any of our results, as shown in Appendix~\ref{sec:dependence_with_enclosed_mass}.}.
This method works well for all galaxies and all times, except for very early epochs when discs are not yet fully formed, and the stellar distributions can present strong asymmetries due to the occurrence of mergers.
For this reason, for times less than $4\,\mathrm{Gyr}$ we instead use the minimum value between the radius enclosing 90\% of the total stellar mass and $f R_{200}(t)$, where $f = R_\mathrm{d}(4~\mathrm{Gyr})/R_{200}(4~\mathrm{Gyr})$ is a normalisation factor, different for each galaxy, such that we get a continuous evolution of $R_{\rm d}(t)$ at $4\,\mathrm{Gyr}$.
The values obtained for $f$ for the galaxy sample are in the range $0.055$--$0.156$, with a mean value of 0.098 and a standard deviation of 0.027.

We apply a similar method to calculate the height of the disc, $h_{\rm d}$, as a function of time.
In this case, in order to evaluate any possible vertical asymmetry in the disc, we consider separately the mass above and below the disc plane, and calculate the vertical distance enclosing 90\% of the stellar mass in the corresponding region\footnote{Note that $h_{\rm d}$ refers to the height above and below the disc plane: the thickness of the disc is $2 h_{\rm d}$.} (integrated for all radii).
As long as there are no strong perturbations in the galaxies, we find that the values for the positive and negative regions in the $z$-direction are similar, indicating a high level of vertical symmetry, and take for $h_{\rm d}(t)$ the mean value between those calculated for $z<0$ and $z>0$.
Note, however, that the distribution of stellar mass can be quite asymmetric since the Auriga discs show considerable flaring \citep{Grand2017}, and warping and bending \citep{Gomez2017} in the outer regions.
As in the case of the disc radius, when the standard calculation fails at times less than $4~\mathrm{Gyr}$, we instead use $h_{\mathrm d}=f R_{200}(t)$, where $f$ is the same factor used for estimating $R_{\mathrm d}$.

Our definition of galactic discs using the stellar mass is based on the fact that one of the aims of the present work is to relate the inflow and outflow rates with the star formation activity and, to this end, using stars is more adequate.
We note, however, that discs can also be defined using the gas component.
In this case, more extended discs are expected because gas density decreases with radius and star formation is less efficient in the outskirts of the discs.
According to \cite{BlandHawthorn2017}, large H\,{\sc i} discs in spiral galaxies are expected; in particular, the authors argue that the radius of H\,{\sc i} discs might extend up to $60~\mathrm{kpc}$ for hydrogen column densities of $N_\mathrm{H} = 10^{18}~\mathrm{cm}^{-2}$.
In our simulations, when taking the cold gas into account (temperatures below $2 \times 10^4~\mathrm{K}$), the discs are, on average, 2--2.3 times larger and 1.3--1.6 thicker compared to the stellar ones.
Cold gas, however, can be found in rotational-support near the galactic centre, but also distributed well within the halo, thus introducing a bias towards larger disc sizes.
On the other hand, star-forming gas (hydrogen number density above $0.13~\mathrm{cm}^{-3}$), is almost completely found near the galactic midplane and might be a better tracer of the stellar disc.
In this case, gas discs have, on average, slightly greater radii and are typically thinner than stellar discs.

\begin{figure*}
    \begin{tabular}{c}
		\includegraphics[trim={.87cm .5cm .03cm .01cm}, clip, scale=.7]{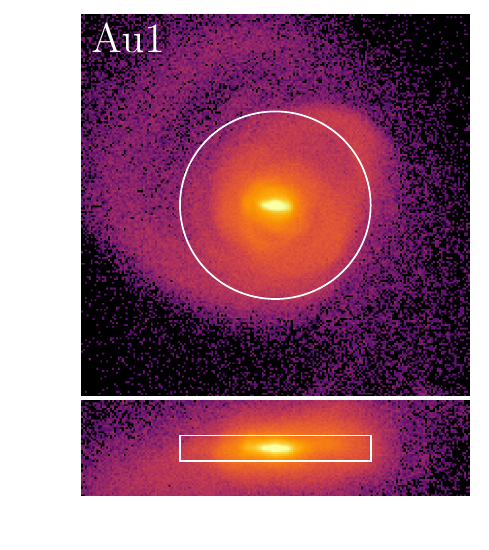}
        \includegraphics[trim={.87cm .5cm .03cm .01cm}, clip, scale=.7]{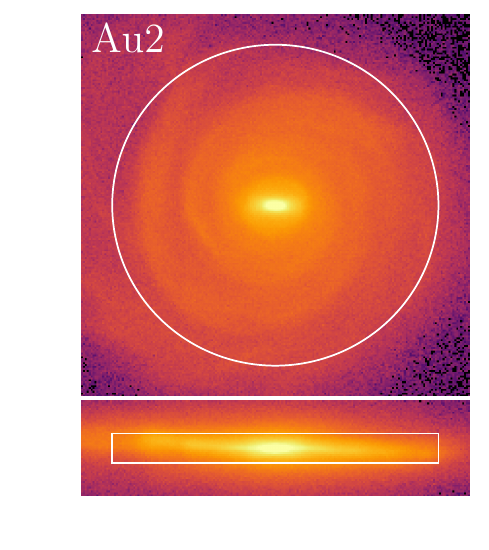}
        \includegraphics[trim={.87cm .5cm .03cm .01cm}, clip, scale=.7]{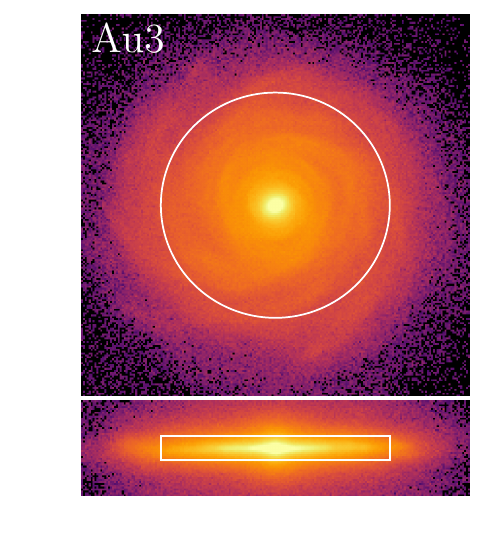}
        \includegraphics[trim={.87cm .5cm .03cm .01cm}, clip, scale=.7]{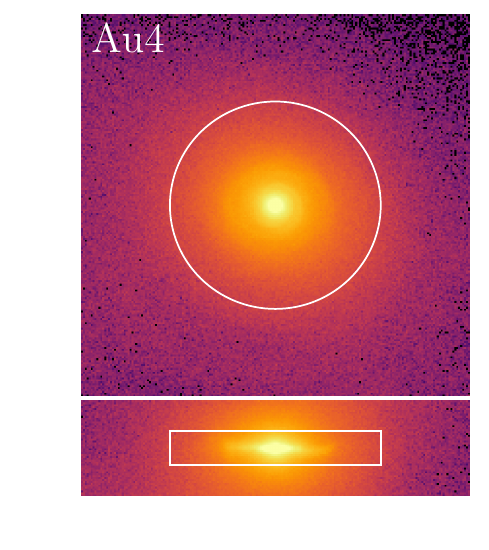}
        \includegraphics[trim={.87cm .5cm .03cm .01cm}, clip, scale=.7]{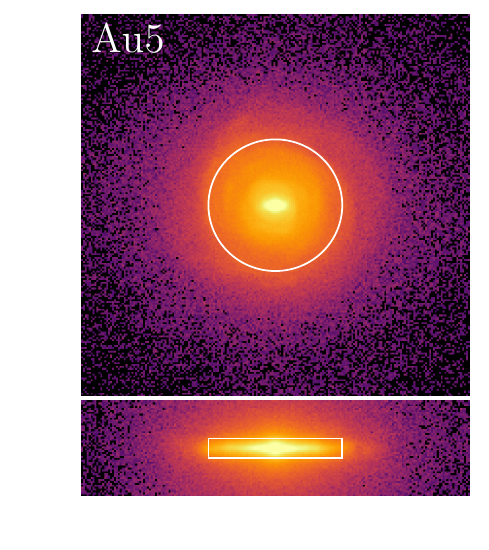}
        \includegraphics[trim={.87cm .5cm .03cm .01cm}, clip, scale=.7]{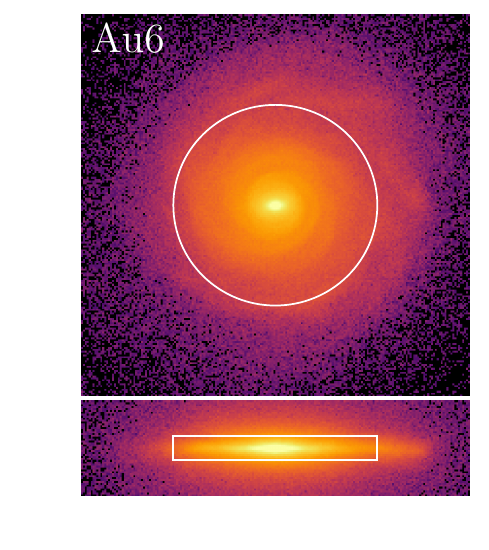} \\
        \includegraphics[trim={.87cm .5cm .03cm .01cm}, clip, scale=.7]{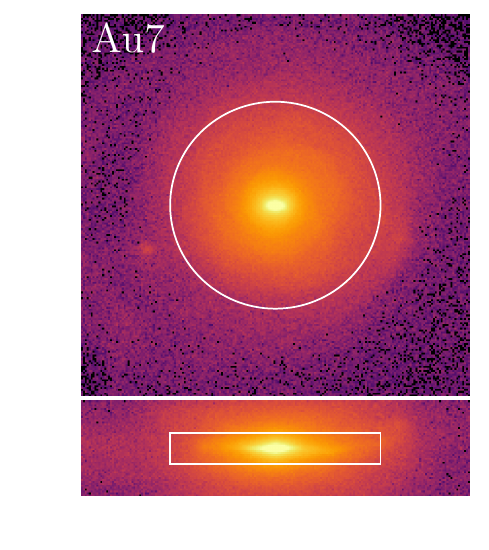}
        \includegraphics[trim={.87cm .5cm .03cm .01cm}, clip, scale=.7]{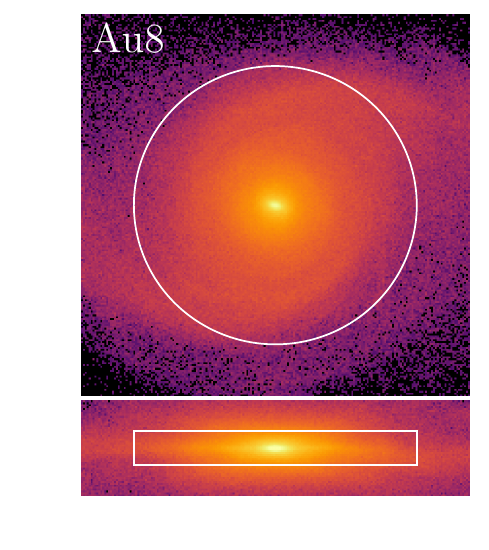}
        \includegraphics[trim={.87cm .5cm .03cm .01cm}, clip, scale=.7]{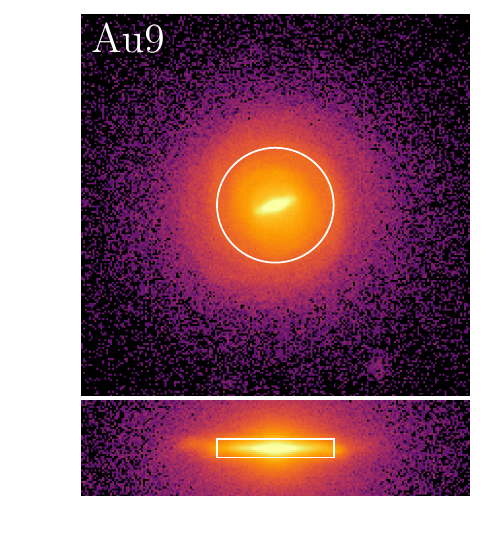}
        \includegraphics[trim={.87cm .5cm .03cm .01cm}, clip, scale=.7]{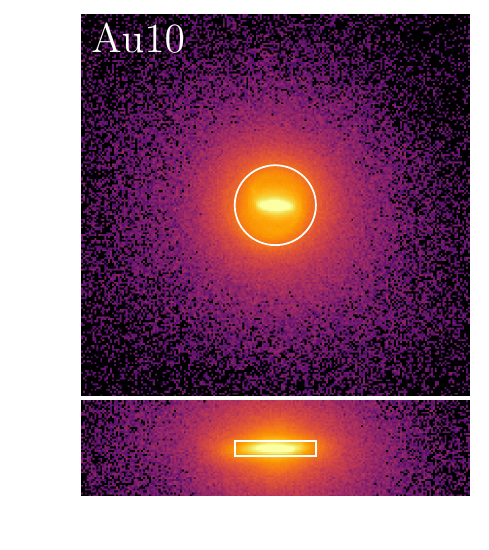}
        \includegraphics[trim={.87cm .5cm .03cm .01cm}, clip, scale=.7]{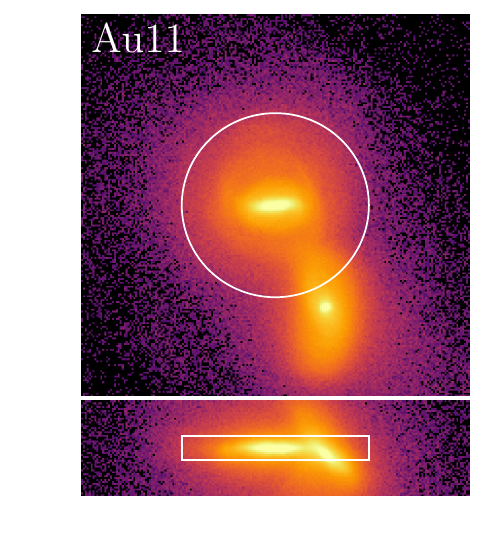}
        \includegraphics[trim={.87cm .5cm .03cm .01cm}, clip, scale=.7]{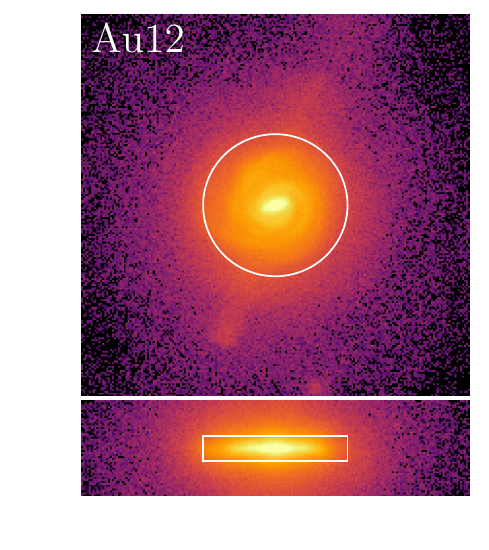} \\
        \includegraphics[trim={.87cm .5cm .03cm .01cm}, clip, scale=.7]{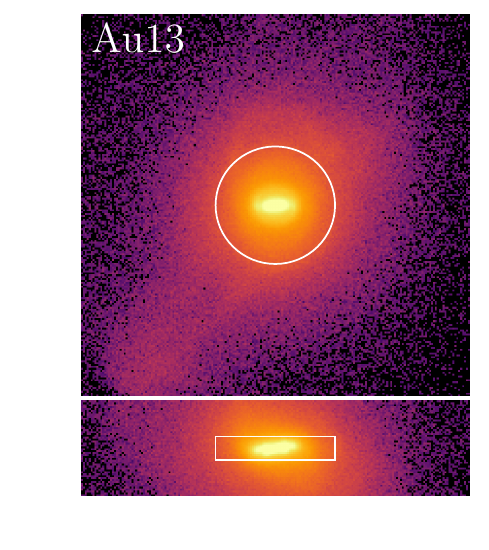}
        \includegraphics[trim={.87cm .5cm .03cm .01cm}, clip, scale=.7]{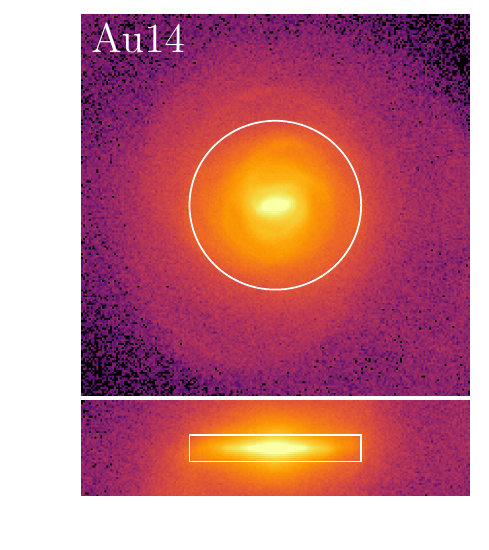}
        \includegraphics[trim={.87cm .5cm .03cm .01cm}, clip, scale=.7]{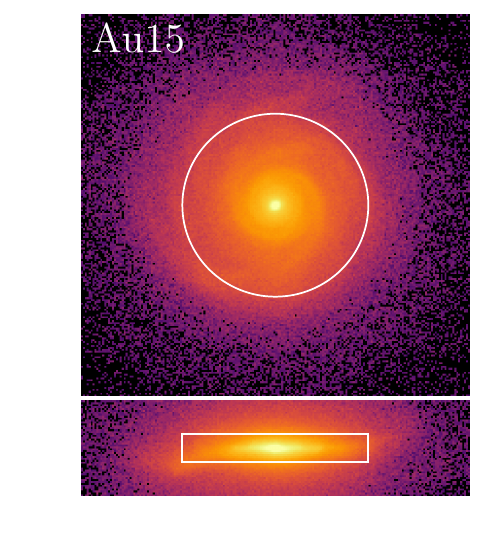}
        \includegraphics[trim={.87cm .5cm .03cm .01cm}, clip, scale=.7]{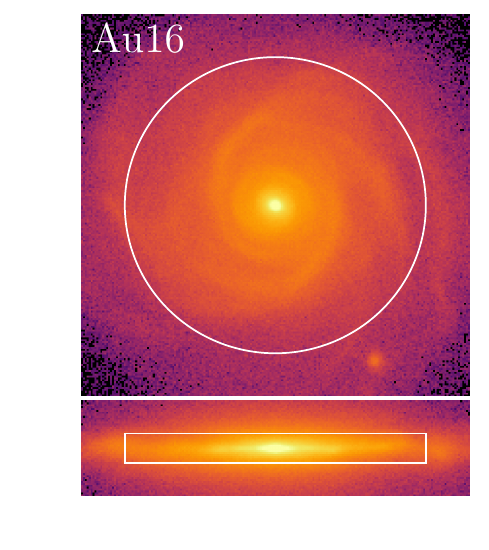}
        \includegraphics[trim={.87cm .5cm .03cm .01cm}, clip, scale=.7]{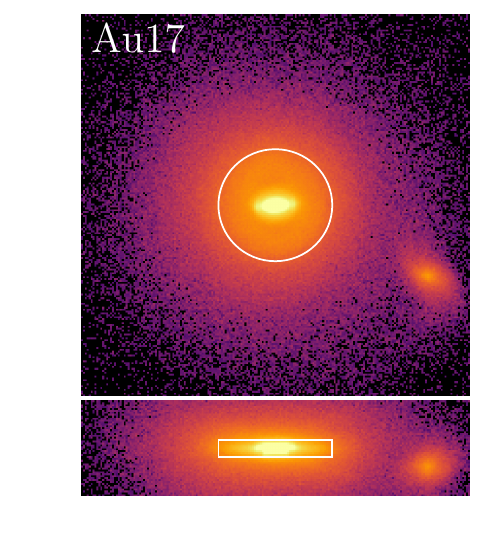}
        \includegraphics[trim={.87cm .5cm .03cm .01cm}, clip, scale=.7]{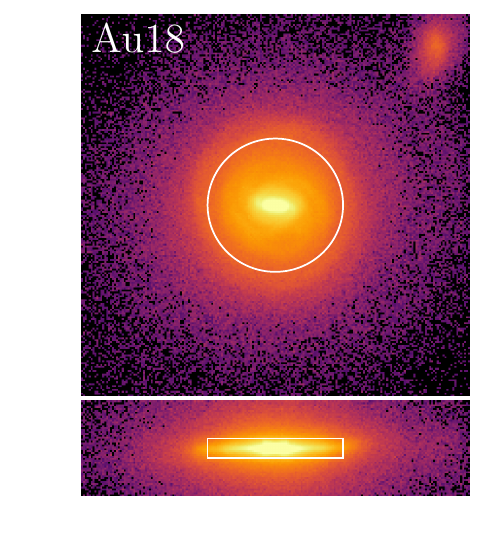} \\
        \includegraphics[trim={.87cm .5cm .03cm .01cm}, clip, scale=.7]{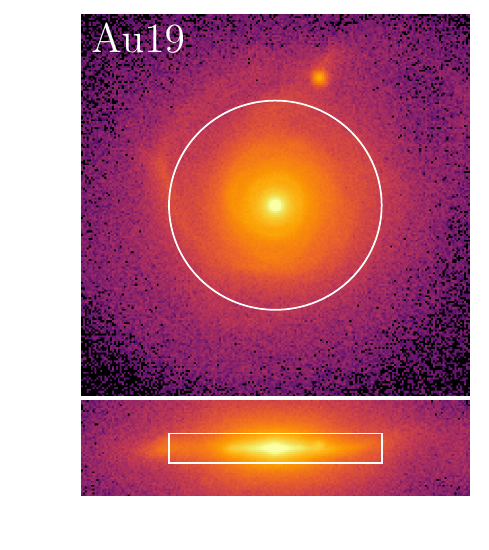}
        \includegraphics[trim={.87cm .5cm .03cm .01cm}, clip, scale=.7]{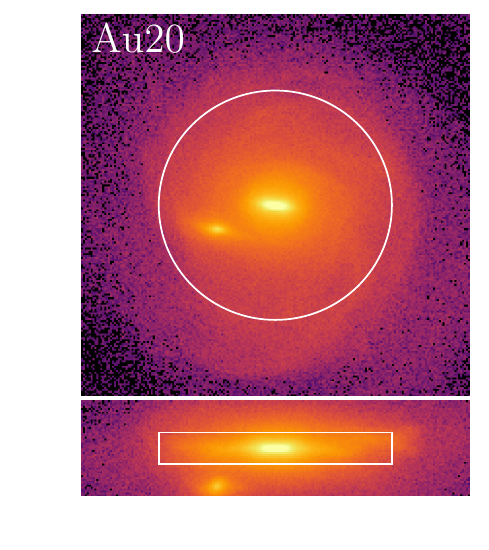}
        \includegraphics[trim={.87cm .5cm .03cm .01cm}, clip, scale=.7]{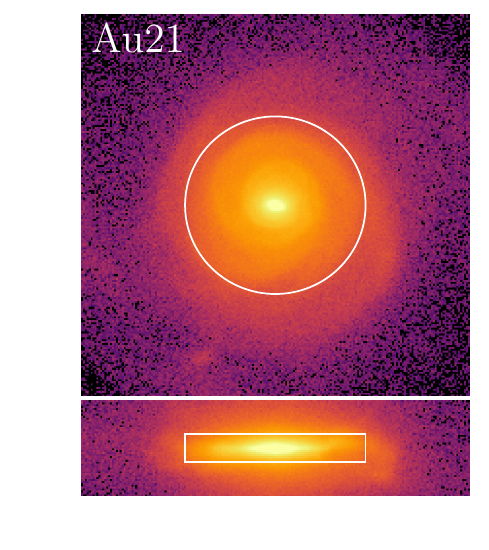}
        \includegraphics[trim={.87cm .5cm .03cm .01cm}, clip, scale=.7]{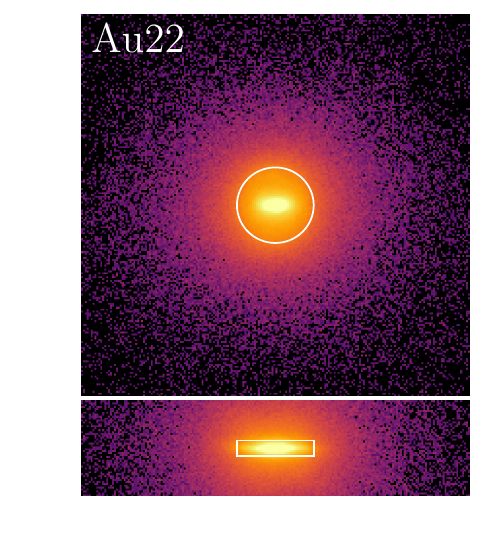}
        \includegraphics[trim={.87cm .5cm .03cm .01cm}, clip, scale=.7]{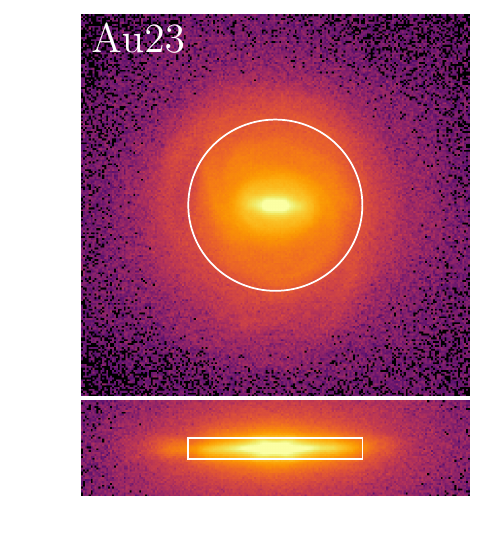}
        \includegraphics[trim={.87cm .5cm .03cm .01cm}, clip, scale=.7]{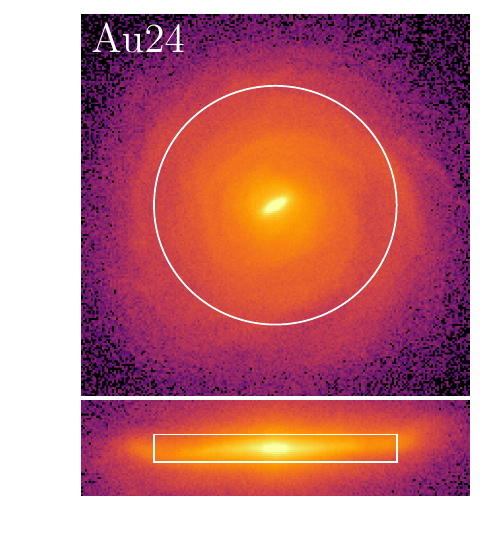} \\
        \includegraphics[trim={.87cm .5cm .03cm .01cm}, clip, scale=.7]{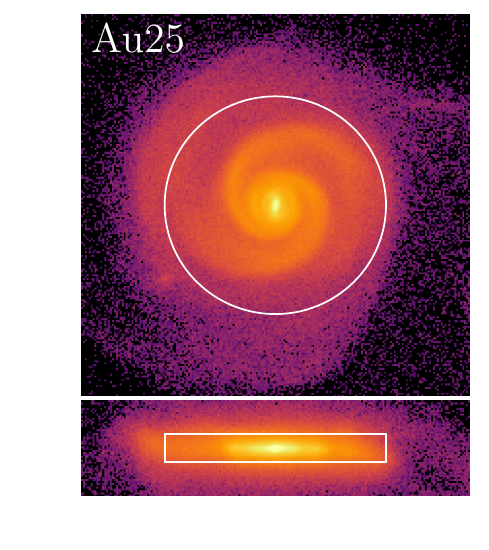}
        \includegraphics[trim={.87cm .5cm .03cm .01cm}, clip, scale=.7]{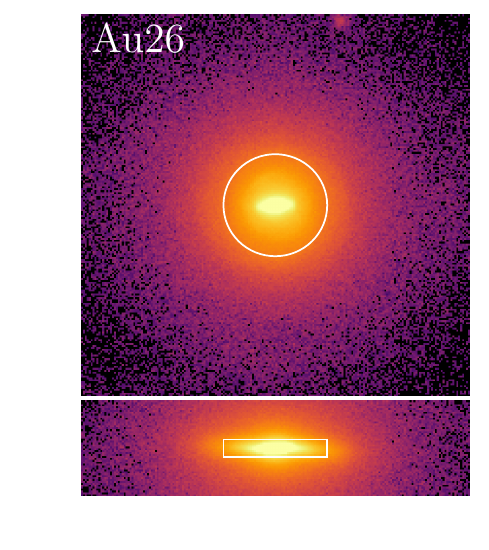}
        \includegraphics[trim={.87cm .5cm .03cm .01cm}, clip, scale=.7]{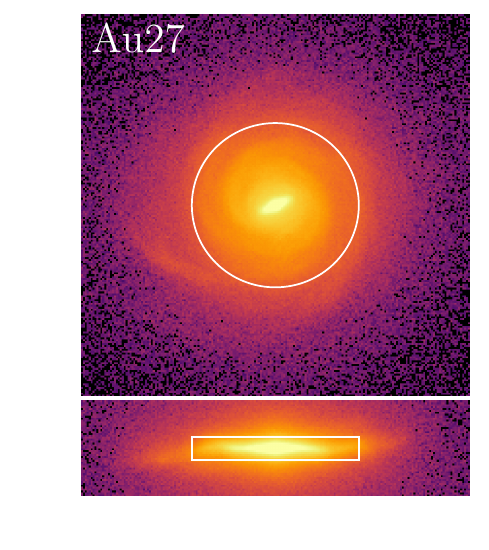}
        \includegraphics[trim={.87cm .5cm .03cm .01cm}, clip, scale=.7]{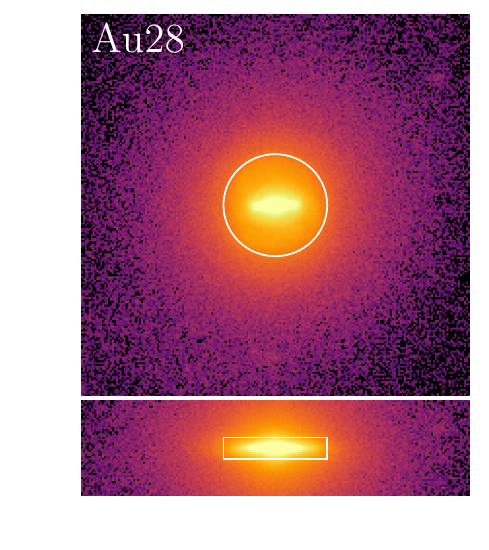}
        \includegraphics[trim={.87cm .5cm .03cm .01cm}, clip, scale=.7]{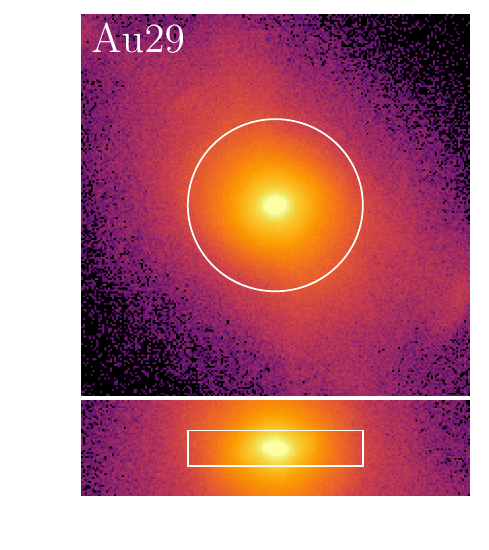}
        \includegraphics[trim={.87cm .5cm .03cm .01cm}, clip, scale=.7]{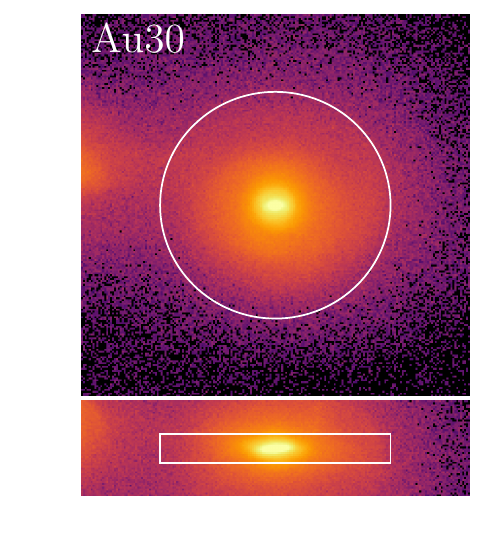} \\
    \end{tabular}
    \caption{Stellar density maps for the Auriga galaxies at $z=0$; for each galaxy we show the face-on view (top) and edge-on view (bottom). The colour map spans five orders of magnitude in projected density using a logarithmic scale. The face-on view shows a cubic region of $80~\mathrm{ckpc}$ on a side and the edge-on view a region of dimensions $80 \times 80 \times 20~\mathrm{ckpc}^3$. The white circles and rectangles indicate the disc region for each view.}
    \label{fig:density_maps}
\end{figure*}

Fig.~\ref{fig:density_maps} shows maps of the projected stellar density distributions of the 30 simulated galaxies at $z=0$, for the face-on and edge-on views, indicating the disc region identified with our method.
The $R_{\rm d}$ and $h_{\rm d}$ values obtained for the galaxies, at $z=0$, are listed in Table~\ref{tab:galactic_properties}.
The present-day disc radii range from $7.9~\mathrm{kpc}$ to $33.7~\mathrm{kpc}$ (consistent with the results presented in \citealt{Grand2017}), while the disc heights range from $1.6~\mathrm{kpc}$ to $3.7~\mathrm{kpc}$.
We note that these values are considerably higher than those corresponding to the thin disc of the MW.
However, this is in part due to our disc height definition\footnote{Note that we do not use disc vertical/radial scale-lengths but rather quantities that can reasonably define the boundaries of the disc.} and also due to disc flaring, as discussed in \cite{Grand2017}.
Significantly lower values for the disc heights are obtained if we consider only the young stars; but in view of the objectives of this work, a less stringent calculation of $h_{\rm d}$ is more adequate.

\begin{figure*}
    \centering
    \includegraphics{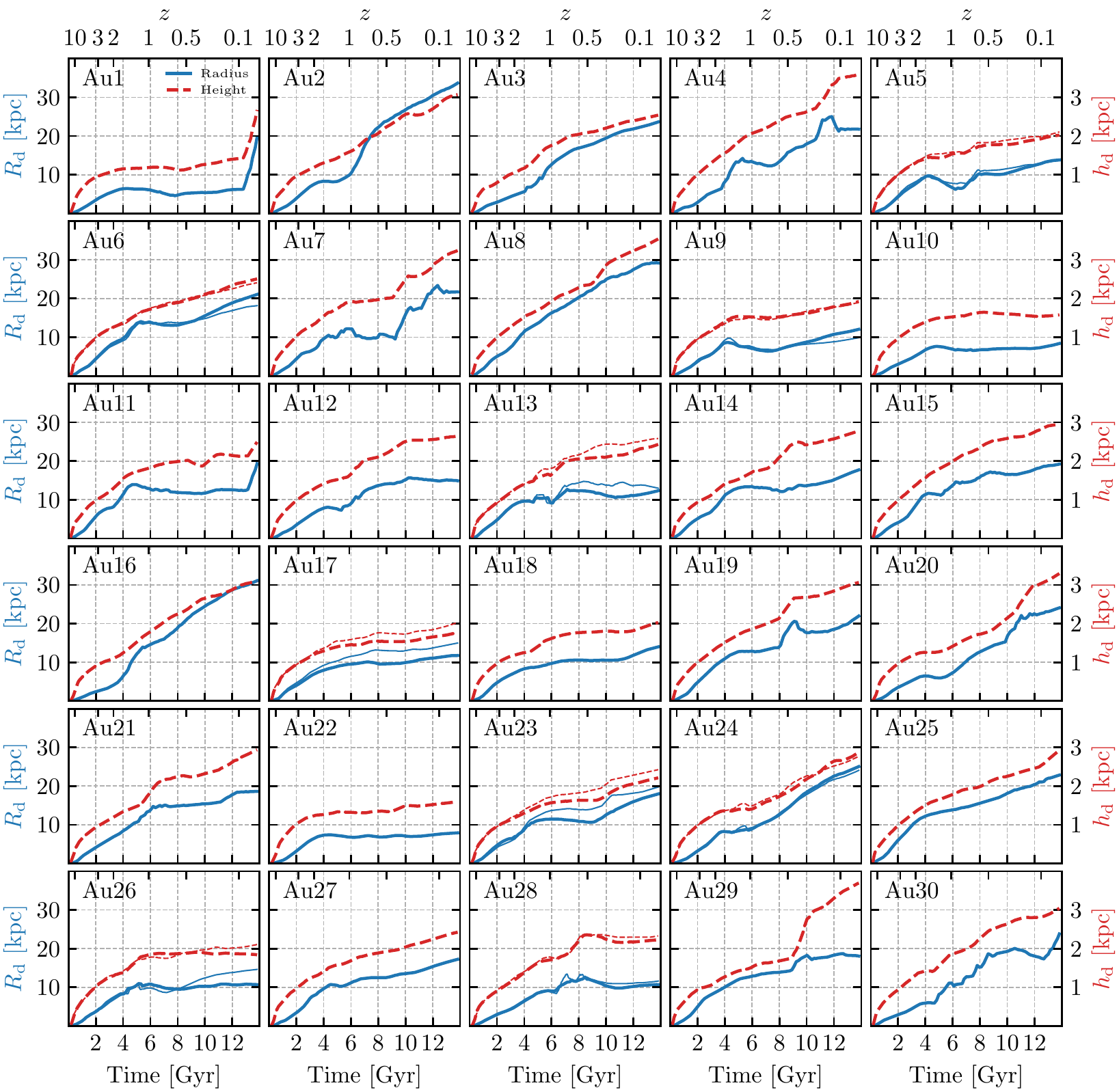}
    \caption{Evolution of the disc radius (blue, continuous line) and height (red, dashed line) in physical coordinates as a function of time for the simulated galaxies. When available, thin lines indicate (using the same colour code) the evolution of these parameters for simulations with tracer particles. For the latter, we apply the same method to calculate the disc parameters; hence, the small differences observed between thin and thick lines are intrinsic to the stellar distribution of each simulation. For clarity, we also add corresponding redshifts in the top panels.}
    \label{fig:disc_sizes}
\end{figure*}

The time evolution of $R_{\rm d}$ and $h_{\rm d}$ for the simulated galaxies is shown in Fig.~\ref{fig:disc_sizes} for all galaxies.
When available, we also include the corresponding evolution for the runs that include tracer particles which, as expected, are in agreement with the standard runs.
From these plots, we observe that, in general, the disc vertical/radial sizes increase with time.
There are, however, significant galaxy-to-galaxy variations, in terms of the disc growth rate and on the typical time periods of maximum growth.

\begin{figure}
    \centering
    \includegraphics[scale=1.2]{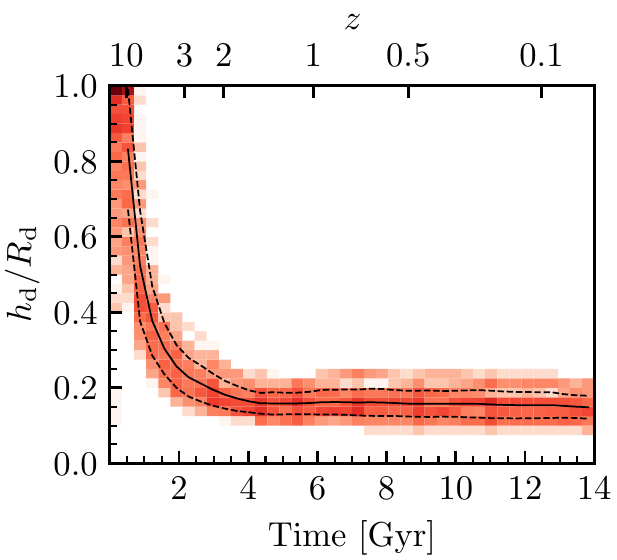}
    \caption{
    Temporal evolution of the ratio $h_\mathrm{d}/R_\mathrm{d}$ for the whole galaxy sample distribution, where redder colours indicate higher counts.
    The black solid line indicates the average of the distribution, and the dashed lines the $\pm\sigma$ standard deviation.
    After $\sim3~\mathrm{Gyr}$ the ratios stay approximately constant with values $\lesssim 0.2$.
    }
    \label{fig:hd_rd_ratio}
\end{figure}

It is worth noting that the very early phases of evolution are characteristic of the formation of the bulge component; and that a disc-like configuration is only present approximately from $2~\mathrm{Gyr}$ on.
Fig.~\ref{fig:hd_rd_ratio} shows that, on average, the ratio between the disc height and the disc radius of the simulated galaxies is $\lesssim 0.2$ from $\sim 3~\mathrm{Gyr}$, staying approximately constant thereafter.

In order to better understand the differences in the disc evolution of the various simulated galaxies, we have estimated the disc-to-total fractions as a function of time.
Note that the disc growth and evolution can be significantly affected along cosmic time due to various mechanisms, such as mergers, interactions and misaligned gas accretion \citep{Scannapieco2009}.
To estimate the disc-to-total mass fractions, we first calculate the circularity distribution of the stars for the various galaxies and for the different times.
The circularity parameter, $\epsilon$, is defined for each star as $\epsilon = j_z/j_\mathrm{circ}$, where $j_z$ is the specific angular momentum in the $z$-direction and $j_\mathrm{circ}(r) = r \sqrt{(GM(r)/r)}$ the specific angular momentum expected for a circular orbit at the star's radius $r$ \citep{Scannapieco2009}.
The circularity distribution for a disc-bulge system is usually characterised by two peaks, a first one around $\epsilon \sim 0$ associated to the bulge (particles with random motion and no net rotation), and a second peak at $\epsilon \sim 1$ associated to a disc-like structure in rotational support.
From the circularity distribution, we calculate the disc-to-total mass fraction, D/T, as $(M_{\epsilon >0} - M_{\epsilon <0})/M_\star$, where $M_{\epsilon >0}$ and $M_{\epsilon <0}$ are the total stellar masses of particles with $\epsilon > 0$ and $\epsilon < 0$, respectively, and $M_\star$ is the total stellar mass.
We note that this definition provides a good estimator for the disc prominence provided the bulge component is non-rotating, which is the case for the simulated galaxies during most of their evolution.

\begin{figure*}
    \centering
    \includegraphics{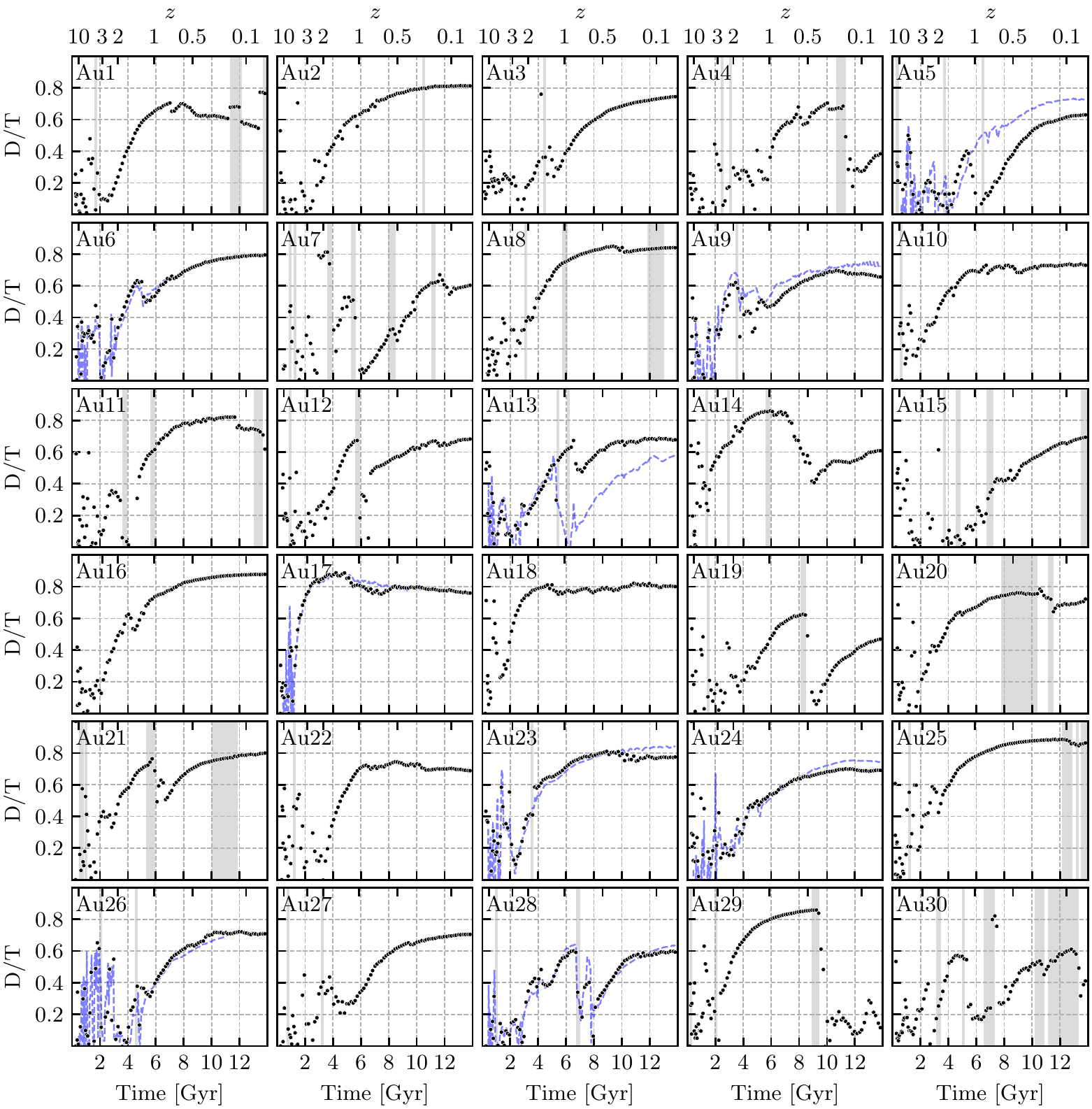}
    \caption{
    Evolution of the disc-to-total mass fraction (D/T) as a function of time for the simulated galaxies (black dots).
    We also show (in light-blue dashed curves) the D/T mass fraction for simulations with tracer particles when available.
    Since the orbit of a satellite can have serious consequences on the evolution of the discs, the shaded background regions indicate the presence of a satellite inside $R_{200}$ with $f_\mathrm{sat} = \frac{M_\mathrm{sat}}{M_\mathrm{cen}} > 0.1$.
    Although most galaxies present stable, long-lasting discs, and show high disc-to-total fractions at $z=0$, there is a considerable variation in the evolution of D/T mainly due to the accretion of satellites and their interaction with disc stars.
    }
    \label{fig:disc_to_total}
\end{figure*}

Fig.~\ref{fig:disc_to_total} shows the evolution of the disc-to-total mass fraction for the 30 simulations.
Most galaxies present stable, long-lasting discs, reaching $\mathrm{D/T} \gtrsim 0.3$ between $\sim 2$ and $4~\mathrm{Gyr}$ (note that 0.3 is a reasonable threshold to consider that discs are well-formed when kinematic estimations are considered, e.g. \citealt{Scannapieco2010}).
We also find galaxies where the discs form very early on (such as Au17 and Au18), as well as discs formed relatively late, most notably in the case of Au15 where a prominent disc forms at $\sim 7~\mathrm{Gyr}$.
Furthermore, we note that although in most cases the discs of our simulated galaxies show a mild, continuous growth in time, we also detect periods of partial/total disc destruction, which are in general followed by a regrowth of the disc, such as in Au14, Au19 and Au28.

In Table~\ref{tab:galactic_properties} we show the obtained D/T values, at $z=0$, for the simulated galaxies.
At the present time, most galaxies have D/T values in the range 0.5--0.9, despite some particular cases such as Au4, Au19, Au29 and Au30 which can be explained by disc instabilities produced by merger episodes.
This can be seen from the background shading in Fig.~\ref{fig:disc_to_total}, where we indicate those times where massive satellites -- with a mass fraction larger than 0.1 -- are identified within the virial radius of the simulated galaxies.

\begin{figure*}
    \centering
    \includegraphics{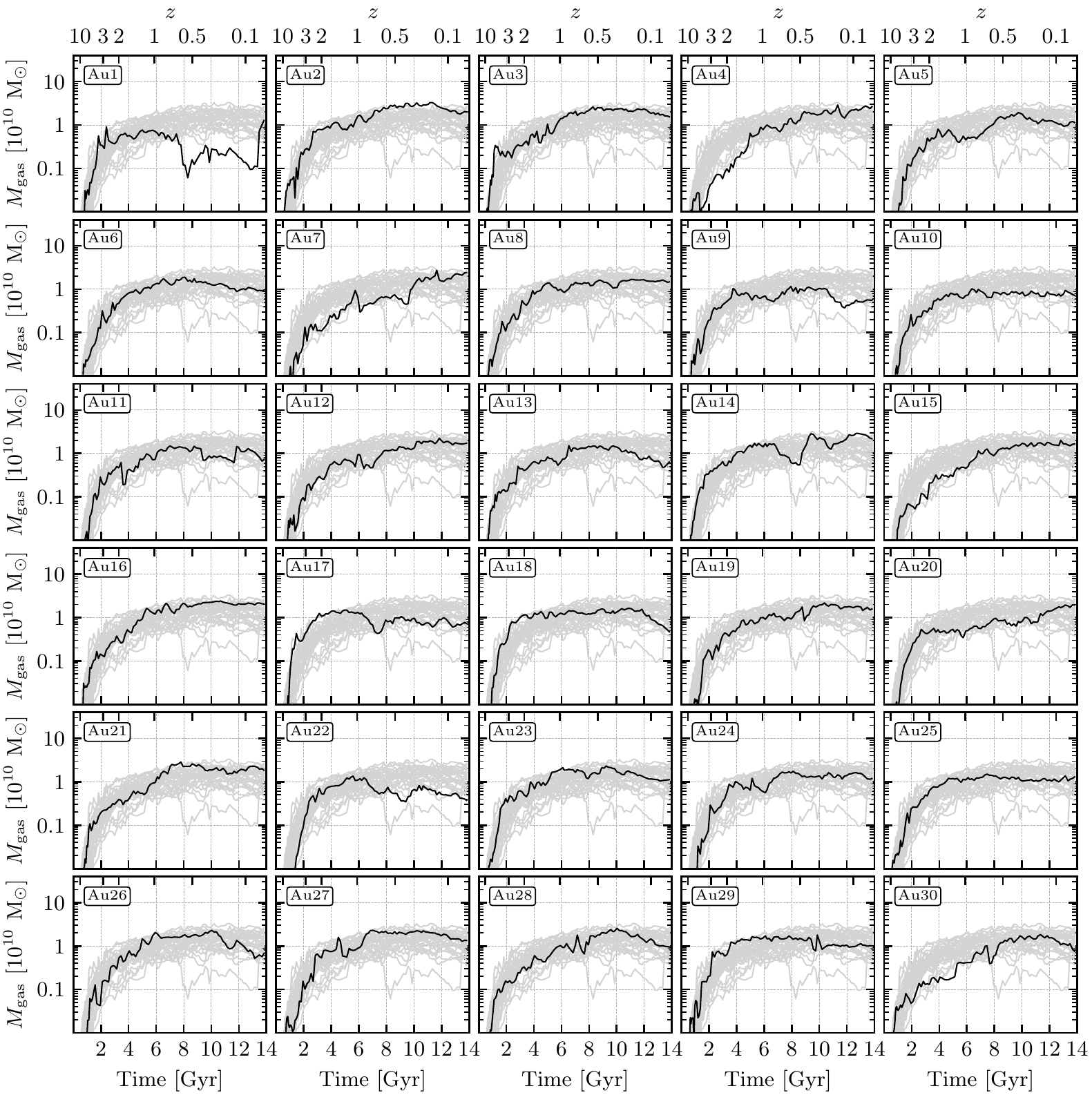}
    \caption{
    Evolution of the gas content in the stellar disc as a function of time for the full Auriga sample.
    The black curve indicates the evolution of the gas mass for the galaxy indicated in each panel; for comparison, the grey background shows curves for the rest of the galaxies.
    The gas mass in the discs is about $\sim10^{10}\,\mathrm{M}_\odot$ at $z=0$ for all galaxies in the sample.
    The evolution, however, can show different behaviours: increasing (e.g. Au4), roughly constant since a few $\mathrm{Gyr}$ (e.g. Au8), or decreasing near the present (e.g. Au18).
    Note, however, that the gaseous and stellar discs need not be aligned during the evolution and, in such cases, the amount of gas mass in the stellar disc is not representative of the gaseous disc's mass.
    }
    \label{fig:disc_gas_mass}
\end{figure*}

With our definitions of the radial and vertical sizes for the discs, we can calculate the amount of gas mass in the disc as a function of time -- which, in turn, comes from the inflowing gas -- which plays a fundamental role in the evolution of the galaxies as it provides the fuel from which the stars form.
Fig..~\ref{fig:disc_gas_mass} shows the evolution of the disc gas mass for the full Auriga sample in a logarithmic scale.
In general terms, the gas reservoir grows rapidly at early times from $\lesssim 10^{8}~\mathrm{M}_\odot$ at $t \lesssim 1~\mathrm{Gyr}$ to $\sim 10^{10}~\mathrm{M}_\odot$ at $\sim 6~\mathrm{Gyr}$.
Afterwards, we find a variety of behaviours: in galaxies like Au3, Au6 and Au13, there is a decrease in the gas reservoir near the present; in galaxies like Au8, Au10 and Au25, the amount of gas in the disc remains roughly constant until $z=0$; in galaxies like Au4, Au7 and Au20, on the other hand, the gas mass keeps increasing until the present.
At $z=0$, the amount of gas in the discs is similar for all galaxies, at $\sim 10^{10}~\mathrm{M}_\odot$.
As we show in the next Section, we see some degree of variety in the net accretion rates, and galaxies can be separated into two groups according to their late-time behaviour: those with decreasing rates and those with constant/increasing ones.
It is possible to appreciate a similar trend in the evolution of the gas mass in the disc region: galaxies with an increasing amount of gas are those included in group G2 (Au4, Au7, Au12, Au15 and Au20) while the ones that show decreasing gas reservoirs near the present are those belonging to group G1 (for example: Au6, Au13 and Au22).
Some of the galaxies in G1, however, show approximately constant gas reservoirs near the present, as is the case of Au10 or Au25 after $\sim 5~\mathrm{Gyr}$, for example.

Finally, it is worth noting that, for the purposes of this work, it is important that the stellar and gaseous discs of the simulated galaxies are aligned, as we are rotating the system using the stellar component.
In principle, misalignment between the stellar and gas discs can occur during galaxy evolution, particularly during mergers/interactions or periods of gas accretion whose angular momentum is not aligned with that of the stellar disc \citep{Scannapieco2009}.
We have checked that most simulated galaxies experience such behaviour during short periods of time; however, the strongest misalignments occur at early times or following merger events, and do not affect our results in any significant way.

\subsection{Computing gas accretion rates and SFR}

In this section we describe our procedure to calculate the net, inflow and outflow gas accretion rates onto the disc region of the simulated galaxies.
As explained above, for the standard Auriga simulations the gas trajectories can not be followed in time, and therefore it is only possible to estimate net accretion rates, whereas the simulations with tracer particles do allow a separate calculation for the inflow and outflow rates.

\subsubsection{Inflow and outflow gas rates}
\label{sec:inflow_and_outflow_rates_calculation}

In the simulations with tracer particles, the inflow rate is calculated as the gas mass per unit time that enters the disc region, and the outflow rate as the equivalent quantity for material leaving the disc at any given time.
In practice, the inflow rate between snapshots $i-1$ and $i$ is the total mass of tracer particles which were outside the disc in snapshot $i-1$ and inside this disc in snapshot $i$, divided by the corresponding time interval.
In order to avoid errors that might be present if the disc size changes significantly between snapshots $i-1$ and $i$, we use the latter for our calculation.
Note that the inflow rate also considers material accreted as gas that was rapidly turned into stars, by taking into account, at snapshot $i$, tracer particles that are either in a gas cell or locked inside a star particle.
A similar method is done to obtain the outflow rate, using particles which moved out from the disc between consecutive snapshots.

Using the inflow and outflow rates obtained, we calculate the net accretion rate as their difference, i.e.,
\begin{equation} \label{eq:net_acc_tracers}
    \dot{M}_\mathrm{net} = \dot{M}_\mathrm{inflows} - \dot{M}_\mathrm{outflows},
\end{equation}
which, by definition, indicates net inflows if $\dot{M}_\mathrm{net}>0$ and net outflows otherwise.

\subsubsection{Net accretion rates}
\label{sec:calc_net_acc}

For the whole Auriga sample, the inflow and outflow rates can not be calculated; however, it is possible to estimate the net accretion rate using the information of the gas cells.
The net accretion rate between 
snapshots $i-1$ and $i$ is computed as:
\begin{equation} \label{eq:net_acc_cells}
    \dot{M}_\mathrm{net} (i) = \frac{M_\mathrm{gas}(i) - M_\mathrm{gas}(i-1) + M_\star}{t(i) - t(i-1)},
\end{equation}
where $M_\mathrm{gas}(i)$ denotes the total mass of gas cells in the disc in snapshot $i$, $t(i)$ is the corresponding time, and $M_\star$ is the mass of stars born in the disc in the considered time-interval.
In order to avoid incorrect results due to the fact that the disc size changes between snapshots (although the changes are in general smooth), we consider in this calculation the disc radius and height corresponding to snapshot $i$.
It is worth noting that, with our definition, we consider the accretion of gas onto the disc from all directions, so the values reported in this work include both vertical and radial contributions.
It is also important to note that the net accretion rates, by definition, can be either positive or negative, corresponding to inflow-dominated or outflow-dominated times, respectively.

Although the prescription we use to calculate the net accretion rates using the information of the gas cells is only an approximation, the simulations with tracer particles allow testing its validity; in fact, our results show that the approximation yields very similar result to the direct calculation of the net accretion, as discussed in Appendix~\ref{app:net_comparison}.

\subsubsection{Star formation rate}

In order to analyse the relation between the inflowing and outflowing material and the star formation process we calculate, at each cosmic time, the SFR in the disc region.
To do this, we take all the stars residing in the disc and compute, for a given snapshot, the amount of mass that was born after the previous one.
Then, to calculate the mass rate, we divide the latter by the time elapsed between the two snapshots considered.

\section{Evolution of gas flows and star formation}
\label{sec:results}

\begin{figure*}
    \centering
    \includegraphics[scale=1.3]{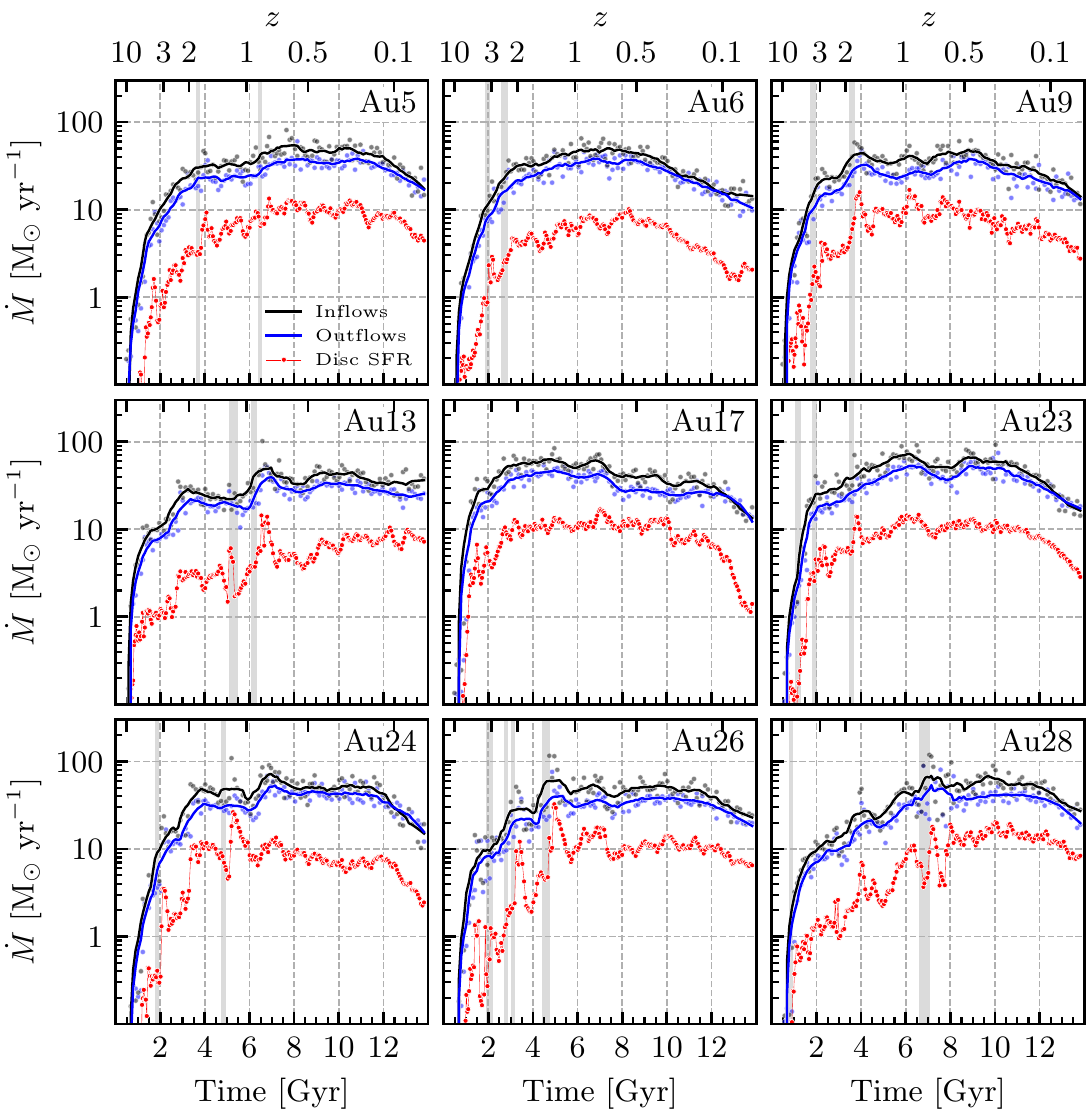}
    \caption{Inflow (black) and outflow (blue) rates calculated with tracer particles for the galaxies that have been re-simulated; lines show the trend and dots the raw data.
    We also indicate the evolution of the SFR in the disc region (red dots).
    Background shading indicates times when there is a satellite inside $R_{200}$ with $f_\mathrm{sat} = \frac{M_\mathrm{sat}}{M_\mathrm{cen}} > 0.1$.
    In general, rates show a rapid increase before reaching a maximum and then decrease to present-day values in the range 10--$40~\mathrm{M}_\odot \, \mathrm{yr}^{-1}$.
    Also note that all rates (inflow, outflow and star-formation) follow roughly the same behaviour.
    }
    \label{fig:inflows_accretion_tracers_reruns}
\end{figure*}

\subsection{Inflowing and outflowing gas rates}
\label{sec:inflow_and_outflow_rates}

In this section, we discuss our results for the inflow and outflow rates onto the simulated discs obtained using the 9 simulations with tracer particles (Au5, Au6, Au9, Au13, Au17, Au23, Au24, Au26 and Au28) after applying the methods described in Section~\ref{sec:inflow_and_outflow_rates_calculation}.

The grey dots in Fig.~\ref{fig:inflows_accretion_tracers_reruns} show the inflow rates onto the discs as a function of time for the 9 simulations; for clarity, we also include a smoothing of the data points as a black line.
The inflow rates show a rapid increase at times $\leq 2~\mathrm{Gyr}$, going from $\sim 0.1~\mathrm{M}_\odot \, \mathrm{yr}^{-1}$ to values of the order of $\sim 20~\mathrm{M}_\odot \, \mathrm{yr}^{-1}$ depending on the galaxy.
These times are characteristic of the collapse and formation of the haloes, when the baryonic distributions evolve from a spheroidal to a disc-like structure (see Fig.~\ref{fig:hd_rd_ratio}).
At intermediate times, between $2~\mathrm{Gyr}$ and $\sim 6~\mathrm{Gyr}$, the inflow rates are in general still increasing.
The late evolution, i.e. after 6--$8~\mathrm{Gyr}$, is characterised by smoothly decreasing inflow rates.
At the present day, the inflow rates integrated onto the simulated discs are in the range 10--$40~\mathrm{M}_\odot \, \mathrm{yr}^{-1}$.
Note that the inflow rates, according to our definition, do not differentiate material reaching the disc for the first time from gas participating in galactic fountains; \cite{Grand2019} found a median recycling time of $\sim 500~\mathrm{Myr}$ in the Auriga haloes, which is greater than the time spacing of the snapshots by an order of magnitude and therefore contribute to our accretion rates.

\begin{figure*}
    \centering
    \includegraphics[scale=1.3]{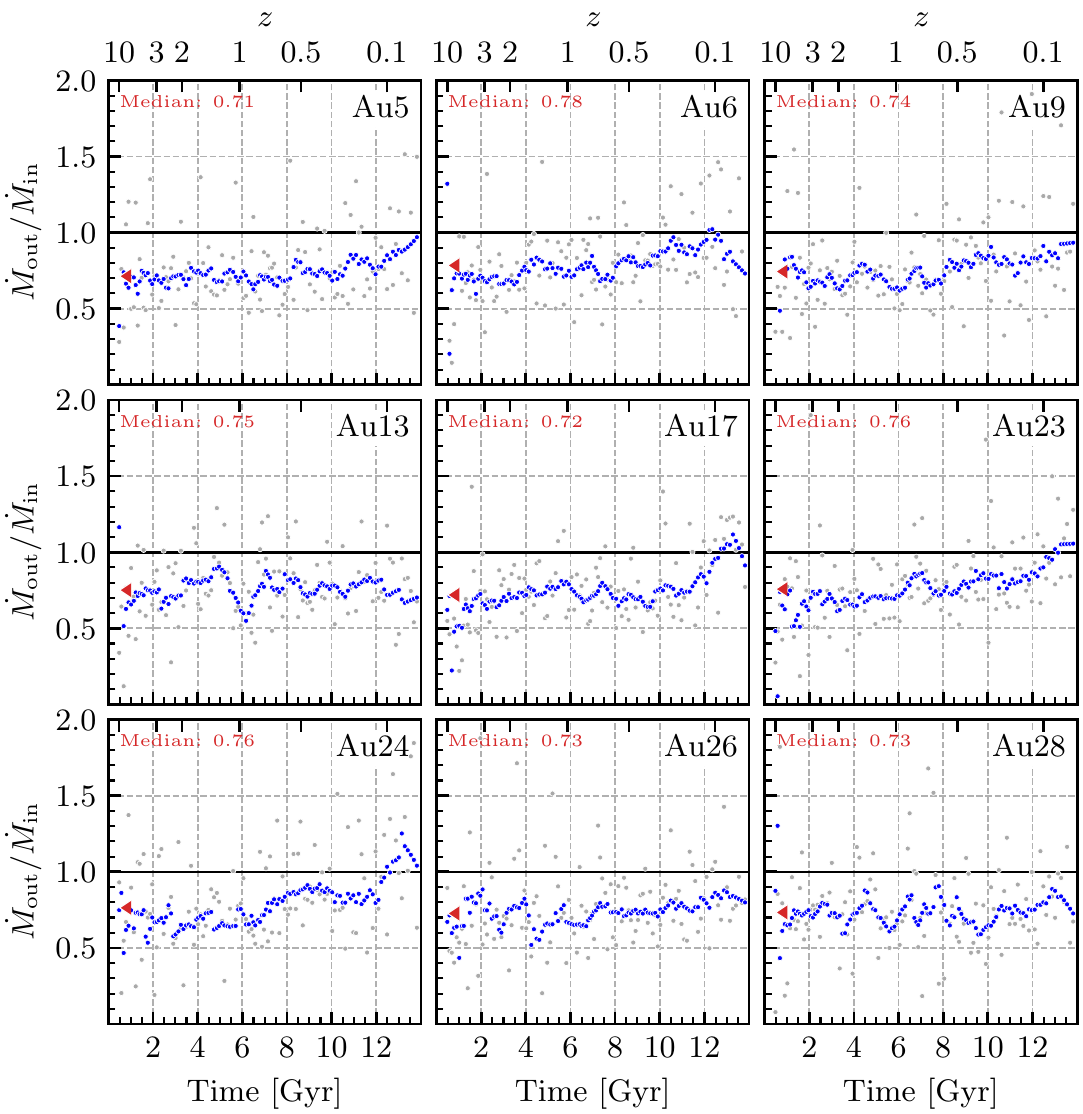}
    \caption{
    Ratio between the outflow and the inflow rate for the raw data of each galaxy (grey dots).
    To guide the eye, blue dots indicate the same ratio but for the smoothed curves (trend) shown in Fig.~\ref{fig:inflows_accretion_tracers_reruns}.
    The red triangle, on the other hand, marks the median value indicated in the top left corner.
    Ratios are typically in the range 0.5--1 for all galaxies with no significant variation over time.
    }
    \label{fig:outflow_inflow_ratio}
\end{figure*}

The outflow rates (blue dots and lines) obtained for the simulated galaxies show a similar behaviour to the inflow rates, but appear systematically at lower values, with present-day values of the order of 10--$30~{\mathrm M}_\odot \mathrm{yr}^{-1}$.
As shown in Fig.~\ref{fig:outflow_inflow_ratio}, the ratio between the outflow and inflow rates are typically in the range 0.5--1 for all galaxies (this is for the smoothed data, but note that more variations are detected for the raw data), with no significant variation over time.
The median values range from 0.71 (for Au5) to 0.78 (for Au6), with an average over all galaxies of $\sim 0.75$.
This means that, on average, the outflowing gas mass is about 25\% smaller than the inflowing mass.

Finally, note that the inflow and outflow rates are not always smooth but present, in many cases, bursts of enhanced inflow/outflow levels.
In general, these can be attributed to interactions and mergers with satellite systems: in Fig.~\ref{fig:inflows_accretion_tracers_reruns}, the grey shades indicate the presence of satellites (with a mass ratio $f_\mathrm{sat} > 0.1$) inside $R_{200}$.
While mergers can produce an increase in the gaseous mass of the discs, provided that the inflowing satellites contain gas, the presence of satellites near the central region can also induce gas inflows leading to a mass increase.
As can be observed in the figure, many of the periods with enhanced rates can be linked to the presence of satellites near the central regions, as most evident in the cases of Au13 at $\sim 6~\mathrm{Gyr}$ and Au26 at $\sim 4.5~\mathrm{Gyr}$.

\subsection{Relation with the SFR}

\begin{figure*}
    \centering
    \includegraphics[scale=1.3]{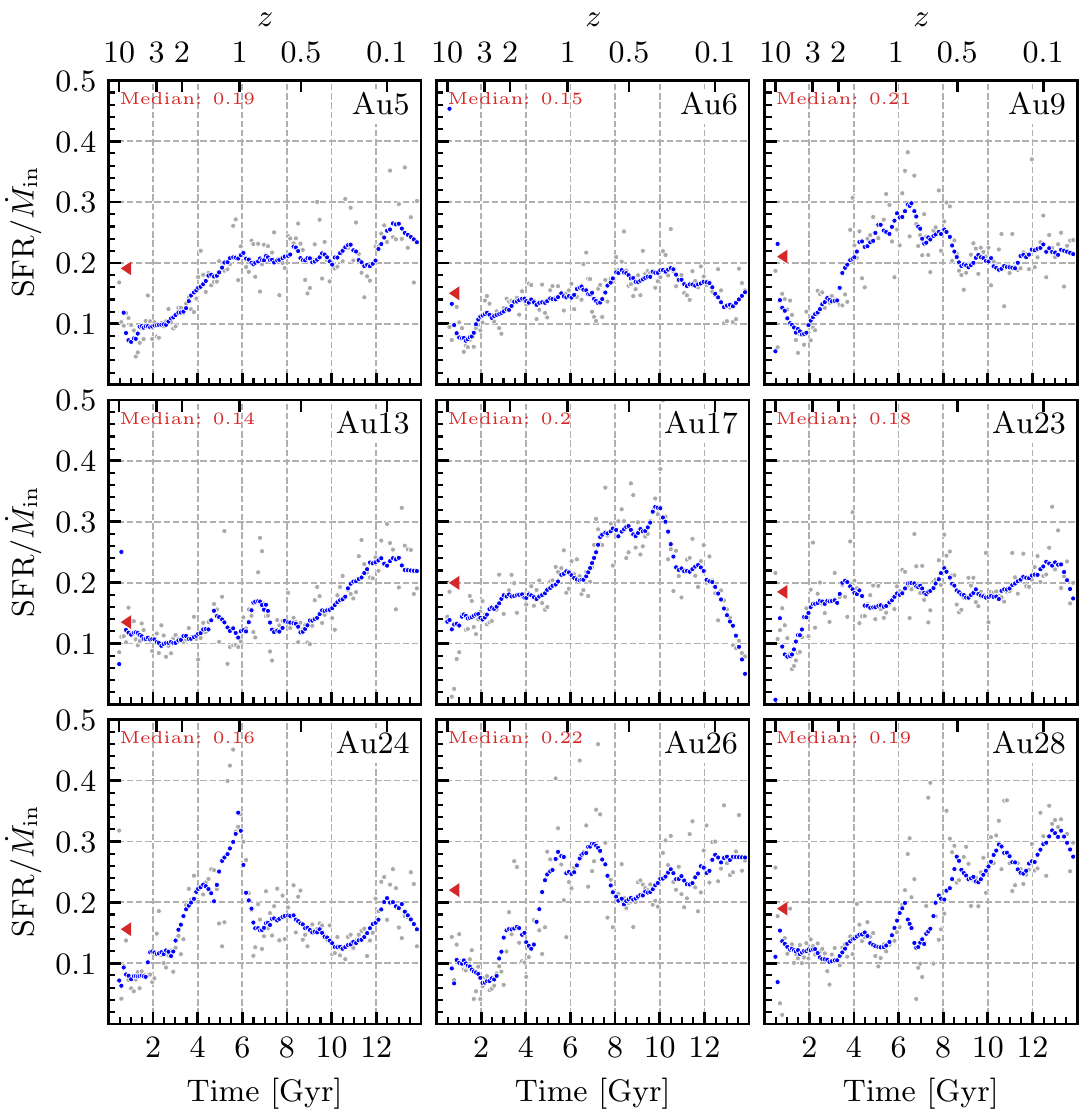}
    \caption{
    Ratio between the SFR and the inflow rate for the raw data of each galaxy (black dots).
    Blue dots indicate the same ratio but for the smoothed curves (trend) shown in Fig.~\ref{fig:inflows_accretion_tracers_reruns}.
    The red triangle marks the median value indicated in the top left corner.
    Ratios vary in the range 0.1--0.3 and are, in general, higher at late times.
    }
    \label{fig:sfr_inflow_ratio}
\end{figure*}

As shown in the previous Section, the evolution of inflow and outflow rates of each simulated galaxy have a similar shape, despite their different values, and their ratios are approximately constant in time.
This is expected, as inflowing and outflowing material are linked through the star formation activity in the discs: while inflow increases the amount of gas mass which can be converted into stars, outflows are a direct consequence of supernova explosions and stellar winds produced rapidly after stars are created.
The SFRs of the simulated galaxies, shown in red in Fig.~\ref{fig:inflows_accretion_tracers_reruns}, therefore establish a link between their instantaneous inflow and outflow rates.

The ratio between the SFR and the inflow rate, as a function of time, is shown in Fig.~\ref{fig:sfr_inflow_ratio} for the simulated galaxies.
In general, $\mathrm{SFR}/\dot{M}_{\mathrm{in}}$ varies in the range 0.1--0.3, with median values for the complete evolution between 0.14 (for Au13) and 0.22 (for Au26) and an average over all galaxies of 0.18.
In general, for the late times characteristic of the formation of the discs, $\mathrm{SFR}/\dot{M}_{\mathrm{in}}$ is higher compared to the values obtained at earlier times.

\begin{figure*}
    \centering
    \includegraphics[scale=1.3]{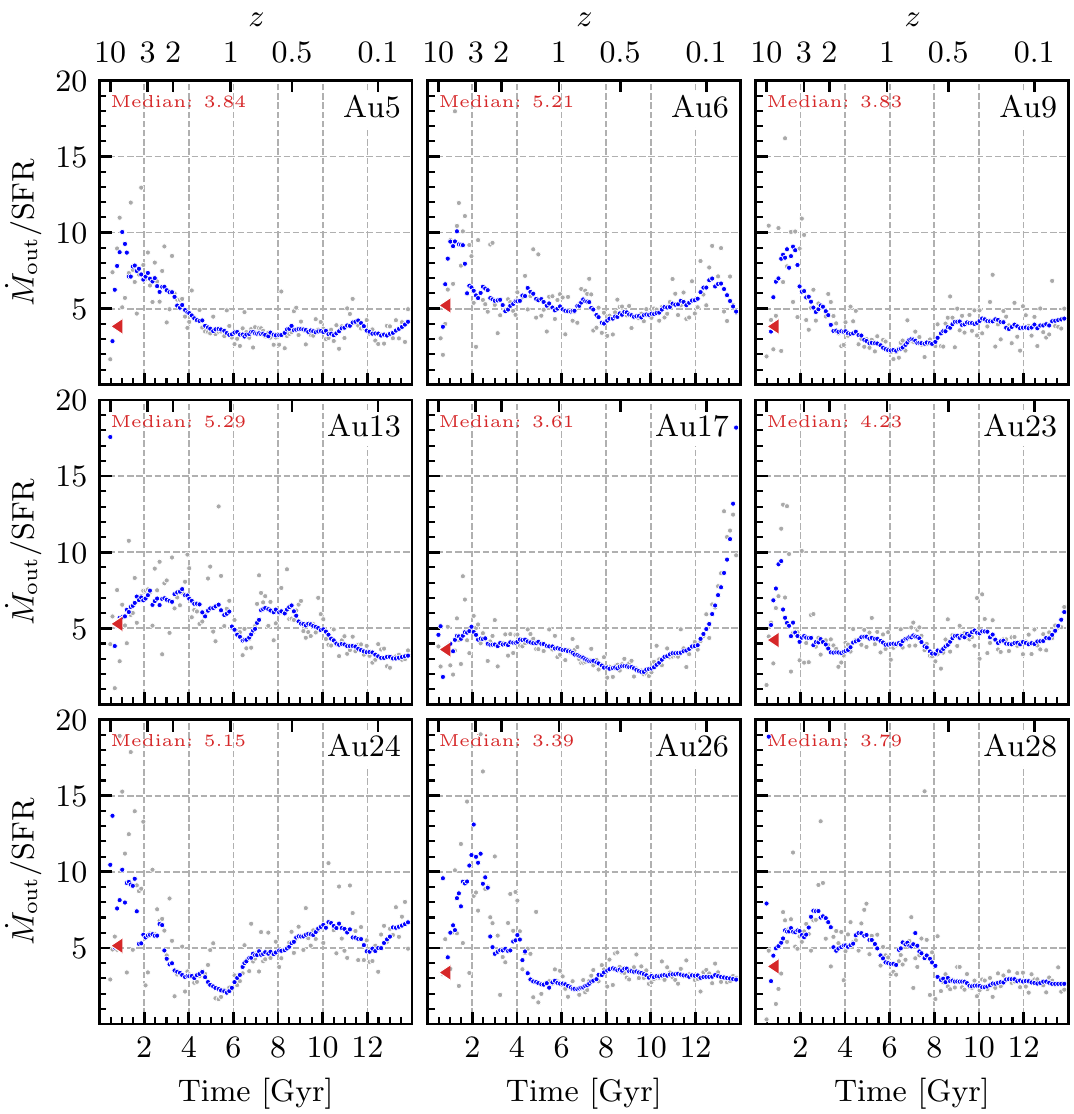}
    \caption{
    Ratio between the outflow rate and the SFR for the raw data of each galaxy (black dots).
    Blue dots indicate the same ratio but for the smoothed curves (trend) shown in Fig.~\ref{fig:inflows_accretion_tracers_reruns}.
    The red triangle marks the median value indicated in the top left corner.
    Ratios vary in the range 3--6 and the median, averaged over the sample of re-simulated galaxies, is 4.26.
    }
    \label{fig:outflow_sfr_ratio}
\end{figure*}

Following the star formation activity in the discs, outflows are mainly produced by gas motions derived from supernova explosions, which provide heat and pressure to the gas around young stars.
As a result, the star formation and outflow rates are related, as can be seen in Fig.~\ref{fig:outflow_sfr_ratio} where we show their ratio as a function of time, i.e. the so-called mass-loading factor.
The $\dot{M}_{\mathrm{out}}/\mathrm{SFR}$ ratios are similar for all galaxies, with median values of the order of 3--6 after the very early phases.
The median values for the whole evolution range from 3.39 (for Au26) to 5.29 (for Au13), with an average over all galaxies of 4.26.
It is worth noting that the outflow rate versus SFR ratio is significantly higher than one, which results from the effective galactic wind model implemented in the simulations.

\begin{figure*}
    \centering
    \includegraphics[scale=1.3]{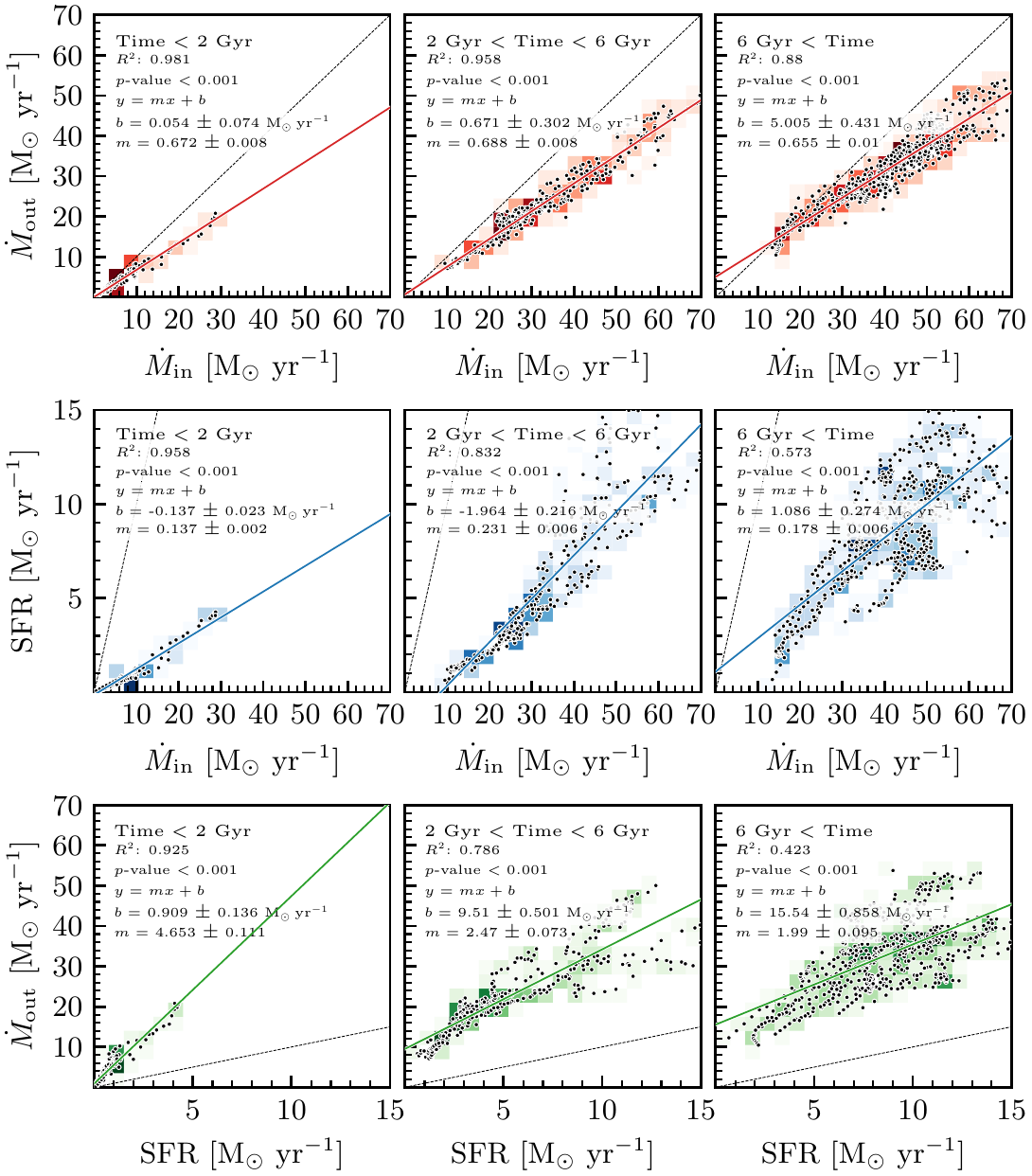}
    \caption{
    Relation between outflow and inflow rates (first row), SFR and inflow rate (second row) and outflow rate and SFR (third row).
    Columns represent different epochs in the evolution: early times (before $2~\mathrm{Gyr}$), intermediate times (from $2~\mathrm{Gyr}$ to $6~\mathrm{Gyr}$), and late times (after $6~\mathrm{Gyr}$).
    Each panel shows data points (black dots); a linear colour density map, and a linear fit (coloured line; best fit parameters are also shown). The three quantities are clearly correlated, with correlation increasing towards early times. For comparison, the identity is included as a dashed line.
    }
    \label{fig:correlations_smoothed}
\end{figure*}

The average behaviour of the relations between inflow, outflow and SFRs can be seen in Fig.~\ref{fig:correlations_smoothed}, separated into the three characteristic time-intervals discussed above.
In this figure, the colours represent the density of the data points (in the smoothed version), the coloured lines correspond to ordinary least squares fits, and the black lines show the one-to-one relation.
It is clear from this plot that the three quantities are correlated, as already evidenced in the previous figures.
A higher dispersion is detected for the interval $t>6~\mathrm{Gyr}$, although this is the longest timescale and a higher dispersion is expected.
The tightest correlation found is for the ratio $\dot{M}_{\mathrm{out}}/\dot{M}_{\mathrm{in}}$, even though the SFR is the link between the two, and the correlations with the SFR show higher dispersions.
These results indicate that the inflow of gas plays a key role in the star formation activity in the discs and in the development of galactic winds.

Furthermore, to confirm that the correlation between inflow and outflow rates is stronger than with the SFR, we applied the jackknife technique (for the whole temporal sample) in order to get an estimate on the error of the $R^2$ statistic. 
The latter is done by excluding from the analysis one galaxy at a time and performing the fits with the rest, thus obtaining nine $R^2$ values from which we can calculate an average value and its standard deviation.
This procedure yielded $R^2 = 0.948 \pm 0.003$ for the inflow-outflow correlation, and $R^2 = 0.807 \pm 0.009$ and $0.75 \pm 0.02$ for the inflow-SFR and outflow-SFR correlations, respectively, thus confirming that the inflow and outflow rates are more closely correlated with each other than with the SFR.

\subsection{Net accretion rates}

We now turn our attention to the analysis of the net accretion rates onto the simulated discs, which we calculate for the full Auriga sample.
As discussed in Section~\ref{sec:calc_net_acc}, the net accretion rates can be estimated from the gas cells (Eq.~\ref{eq:net_acc_cells}), as well as directly from the inflow and outflow rates obtained in the simulations with tracer particles (Eq.~\ref{eq:net_acc_tracers}).
In Appendix~\ref{app:net_comparison}, we show that both methods yield similar results for the simulations allowing both calculations, validating the approximation made in the cases where no tracer particles are considered.

\begin{figure*}
    \centering
    \includegraphics{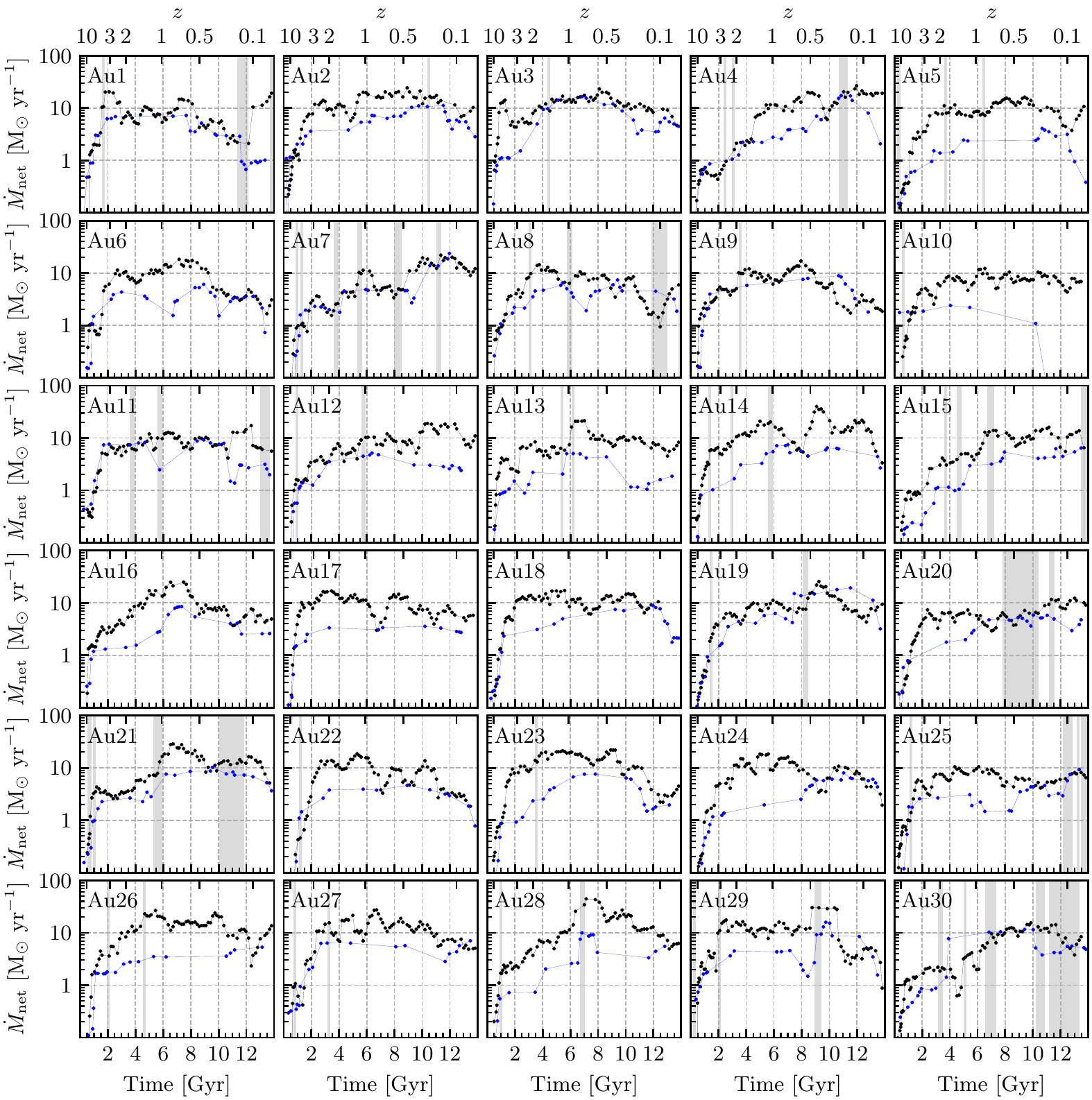}
    \caption{
    Temporal evolution of the net accretion rates for simulated galaxies calculated using cells.
    Black lines indicate positive (inflow-dominated) accretion values, while blue dots indicate negative (outflow-dominated) accretion values (shown in absolute value).
    As before, background shades indicate times when we detect satellites with $f_\mathrm{sat} = \frac{M_\mathrm{sat}}{M_\mathrm{cen}} > 0.1$, inside the corresponding virial radius.
    Although galaxies follow a similar pattern, there are considerable differences in the evolution of the net accretion rate.
    Most galaxies show a decaying rate at late times, but we also observe increasing (e.g. Au4) and approximately constant (e.g. Au10) behaviours.
    }
    \label{fig:net_accretion_cells}
\end{figure*}

The net accretion rates as a function of time for the 30 Auriga galaxies are shown in Fig.~\ref{fig:net_accretion_cells}.
As explained above, the net accretion rates can be either positive or negative, indicating inflow- or outflow-dominated times, respectively.
In this figure, we show separately the results for positive rates (black lines) and negative rates (in absolute value, blue lines).
We find that the net accretion rates are in general positive for all galaxies, with about 70 to 90\% of the data points in this regime.
This result indicates that, in general, there is a net inflow in the disc region of the simulated galaxies, while the occurrence of net outflows is less common.

Fig.~\ref{fig:net_accretion_cells} shows that the net accretion rates for the inflow-dominated times for all galaxies follow a similar pattern, with a rapid increase in the accretion levels at early times, and a smoother evolution at late times, similar to the behaviour of the inflow and outflow rates shown in Section~\ref{sec:inflow_and_outflow_rates}.
Most galaxies show a late-time evolution characterised by decaying accretion rates, reminiscent of an exponential-like behaviour.
However, we also detect galaxies with increasing (e.g. Au4) or approximately constant (e.g. Au10) net accretion rates at late times.
Note that, in most cases, the net accretion rates reach a maximum between 6 and $8~\mathrm{Gyr}$, with values of the order of $\sim 20 ~\mathrm{M}_\odot \, \mathrm{yr}^{-1}$ for most galaxies, and a few cases, such as Au14 and Au28, achieving higher values of approximately $40 ~\mathrm{M}_\odot \, \mathrm{yr}^{-1}$.

Similarly to our findings for the inflow and outflow rates, sudden changes in net accretion levels are observed, which are in most cases related to mergers and interactions with other systems (as can be seen from the coloured shades included in the figure).
In general, we find that mergers/interactions induce an increase in the net accretion levels during the period of interaction or right after the merger event (see, e.g. the cases of Au1, Au7, Au8, Au21 and Au29).

In the case of the outflow-dominated times, not only we find that only $\sim 10$--$30\%$ of the data points are in this regime, but also that the associated net rates are significantly smaller compared to those obtained for the inflow-dominated times.
Our simulations therefore show that, in the disc region, net inflows of gas are more common and important than net outflows, consistent with the findings of Section~\ref{sec:inflow_and_outflow_rates} and in line with the observations of \cite{Fox2019} for the MW.

\section{Temporal evolution of gas flows in MW analogues}
\label{sec:MWanalogs}

The previous section focused on the relation between the different gas flow rates for the whole Auriga galaxy sample. In the following, we analyse the temporal evolution of gas flows in galaxies identified as MW analogues. In particular, we select a subsample of MW-type systems and calculate the average inflow, outflow and net accretion rates onto the disc region as a function of time.
As explained above, the behaviour of the net accretion rates (Fig.~\ref{fig:net_accretion_cells}) naturally separates galaxies into two groups at late times: those with decreasing rates and those with constant/increasing ones.
When examining these groups in detail, we find that they show differences in terms of the evolution and stability of their discs, and in some cases in terms of the disc formation time.
With this information, the galaxies were separated in two groups, which allows us to quantify the average behaviour of systems with a similar evolution.

The first group, referred to as G1, is composed of those galaxies with late-time decreasing accretion rates, i.e. Au2, Au3, Au6, Au8, Au9, Au10, Au11, Au13, Au14, Au16, Au17, Au18, Au21, Au22, Au23, Au24, Au25, Au26 and Au27.
The discs of these galaxies show a smooth growth during, at least, the last $\sim 8~\mathrm{Gyr}$ of evolution, similarly to the MW \citep[e.g.,][]{Minchev2013, Helmi2020}: for these reasons, we consider this group as a MW analogues sample.
The second group, referred to as G2, includes Au4, Au7, Au12, Au15 and Au20, being composed of galaxies with late-time increasing (or approximately constant) rates.
All these galaxies experienced episodes of partial/total destruction of their discs after $4~\mathrm{Gyr}$, related to merger events, except for Au15, which forms its stellar disc relatively late.
Finally, galaxies that exhibit strong perturbations in the D/T evolution or whose accretion rates are highly irregular were excluded from both groups (note that the discs of these galaxies might not be well defined during perturbed times).
These galaxies are Au1, Au5, Au19, Au28, Au29 and Au30.

In the next subsections, we calculate the average net, inflow and outflow rates for the two groups, focusing on G1 which comprises our MW analogues.

\subsection{Net accretion rates} \label{subsec:net_acc}

\begin{figure*}
    \centering
    \hspace*{-0.4cm}
    \includegraphics[scale=1.15]{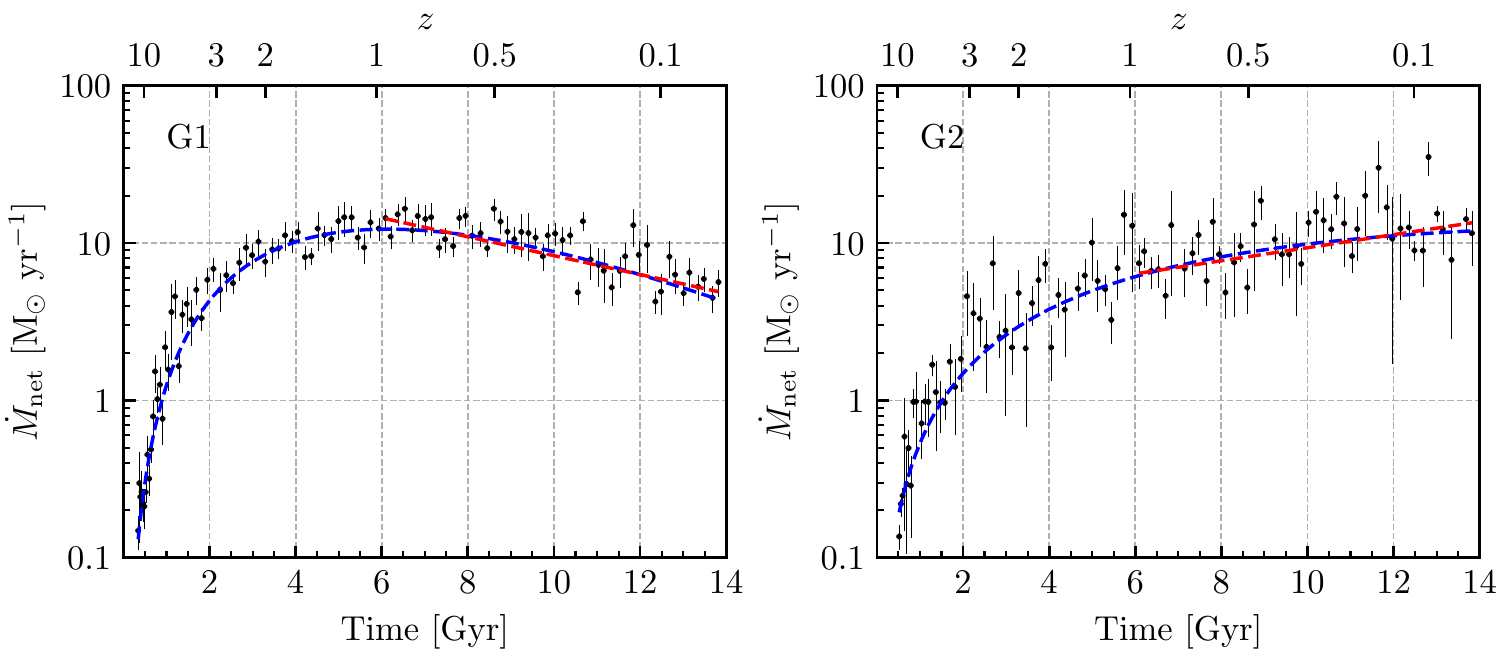}
    \caption{Average net accretion rate for the two groups of galaxies discussed in the text. G1 is comprised of ``well-behaved'' galaxies while G2 includes those that present an increasing accretion rate up to the present. Some galaxies (Au1, Au5, Au19, Au28, Au29 and Au30) were excluded from the groups due to the partial/total destruction of their discs. Error bars indicate plus or minus one standard deviation of the mean. Dashed lines indicate fits, using a Schechter function (blue) and a exponential function (red) starting at $6~\mathrm{Gyr}$.}
    \label{fig:net_accretion_cells_groups}
\end{figure*}

The left-hand panel of Fig.~\ref{fig:net_accretion_cells_groups} shows the average net accretion rates for the MW analogues, together with the $1\sigma$ standard deviation around the mean (indicated as error bars).
These curves have been calculated using the inflow-dominated times which, as explained in the previous section, comprise typically more than 85\% of the data points.
The average net accretion rate shows a rapid increase at early times, reaching a maximum of $\sim 10~\mathrm{M}_\odot \, \mathrm{yr}^{-1}$ at $\sim 6~\mathrm{Gyr}$, and exhibiting an exponential-like behaviour since then.
At $z=0$, the average net accretion rate of this group is $\sim 5~\mathrm{M}_\odot \, \mathrm{yr}^{-1}$.

We also show the average net accretion rate for G2 in the right-hand panel of Fig.~\ref{fig:net_accretion_cells_groups}.
In this case, and following the behaviour of its constitutive galaxies, we see a slower growth of the net accretion rate at early times compared to G1, and increasing net accretion rates until the present time.
We also find higher dispersion levels compared to G1, which is expected because the galaxies in G2 have more dissimilar and less smooth net accretion rates and, moreover, the number of galaxies in this group is lower.

In order to better describe the average net accretion rates of the MW analogues, we performed an exponential fit of the form:
\begin{equation*}
    \dot{M}_\mathrm{net}(t) = A \exp \left( - \frac{t}{t_0} \right),
\end{equation*}
where $t$ denotes the cosmic time, $A$ is the amplitude and $t_0$ the characteristic time-scale of the exponential.
We started the fit at $6~\mathrm{Gyr}$, which corresponds to the maximum average net accretion rate of G1.
We also made a fit in the case of G2, using the same time threshold, in order to better compare the net accretion rates of the two groups\footnote{We have tested that the results of the fits are not affected by the time threshold used to start the fit, as long as we do not consider the very early times which clearly do not follow an exponential-like behaviour.}.
The exponential fits are shown in Fig.~\ref{fig:net_accretion_cells_groups} in dashed-red lines, and the parameters obtained for the fits are shown in Table~\ref{table:fit_parameters_exponential_groups}.

\begin{table}
    \caption{Amplitude and time-scale parameters of the fit using a exponential function for the average net accretion rate of each group. These fits are shown in red in Fig.~\ref{fig:net_accretion_cells_groups}.}
    \centering
    \begin{tabular}{lcc}
    \hline
    Group & $A$ [$\mathrm{M}_\odot\,\mathrm{yr}^{-1}$] & $t_0$ [Gyr] \\
    \hline
    G1		& $33.1 \pm 3.8$ & $7.2 \pm 0.7$ \\ 
    G2		& $3.6 \pm 0.7$ & $-10.5 \pm 1.9$ \\
    \hline
    \end{tabular}
    \label{table:fit_parameters_exponential_groups}
\end{table}

For G1, we find that $t_0 = \left(7.2 \pm 0.7 \right)\,\mathrm{Gyr}$, which is consistent with the result of $\left(7.37 \pm 0.50 \right)\,\,\mathrm{Gyr}$ obtained by \cite{Nuza2019} for galaxy g106r, and to the typical values required by CEMs for the disc (see \citealt{Nuza2019} and references therein).
In the case of G2, the typical time-scale is $\left(-10.5 \pm 1.9 \right)\,\mathrm{Gyr}$, which reflects the late-time behaviour of the net accretion rate of the galaxies in this group.

We also used a Schechter function to fit the average net accretion rates; in this case we are able to fit the whole evolution using three fitting parameters.
The Schechter function has the following functional form:
\begin{equation*}
    \dot{M}_\mathrm{net}(t) = A \left( \frac{t}{\tau} \right)^\alpha \exp \left( - \frac{t}{\tau} \right)
\end{equation*}
where $A$ is an amplitude, $\alpha$ is the power-law exponent and $\tau$ is the characteristic time.
Note that this function has two distinctive phases: at early times ($t \ll \tau \alpha$) it grows following a power-law while at late times ($t \gg \tau \alpha$) it follows an exponential decay.
Also note that the maximum of the function occurs at $t = \alpha \tau$, and $\tau$ -- which is positive by definition -- refers to the characteristic time of the exponential regime of the function.
The Schechter fits for G1 and G2 are shown in blue in Fig.~\ref{fig:net_accretion_cells_groups}, and the corresponding parameters are listed in Table~\ref{table:fit_parameters_schechter_groups_static}.

\begin{table}
    \caption{Amplitude, exponent and time-scale parameters of the fit using a Schechter function for the average net accretion rate of each group. These fits are shown in blue in Fig.~\ref{fig:net_accretion_cells_groups}.}
    \centering
    \begin{tabular}{lccc}
    \hline
    Group & $A$ [$\mathrm{M}_\odot\,\mathrm{yr}^{-1}$] & $\alpha$ & $\tau$ [Gyr] \\
    \hline
    G1		& $17.4 \pm 0.6$ & $2.3 \pm 0.1$ & $2.6 \pm 0.1$ \\ 
    G2		& $29.9 \pm 5.1$ & $1.6 \pm 0.1$ & $11.8 \pm 2.7$ \\
    \hline
    \end{tabular}
    \label{table:fit_parameters_schechter_groups_static}
\end{table}

The values obtained for the best fit parameters of the two groups reflect the differences in their behaviour.
In particular, the time corresponding to the maximum of the curves occurs at $\alpha \tau = 6.0~\mathrm{Gyr}$ for G1, and at $\alpha \tau = 18.9~\mathrm{Gyr}$ for G2, which is larger than the age of the Universe.
This occurs because the net accretion rate of G2 is still increasing at the present time, and the maximum of the function will occur in the future.
The values for the amplitudes obtained for G1 and G2 are similar, of the order of $\sim 20\,\mathrm{M}_\odot\,\mathrm{yr}^{-1}$.
For the typical time-scales, we find $\tau = 2.6~\mathrm{Gyr}$ for G1 and $\tau = 11.8~\mathrm{Gyr}$ for G2.

It is worth noting that the time parameter obtained with the Schechter function, $\tau$, has a different meaning than $t_0$, the typical time-scale of the exponential fit.
The first one is a measure of the decay of the exponential regime of the Schechter function at times $t \gg \alpha \tau$, while $t_0$ represents the time-scale of the exponential behaviour between the minimum and maximum times of the fit (in our case, times between $\sim 6~\mathrm{Gyr}$ and the present).

Finally, let us note that Fig.~\ref{fig:net_accretion_cells_groups} has been constructed without any renormalisation, even though the mass of the systems is slightly different.
We have checked that normalising with the virial mass does not affect any of our results in terms of the fitting parameters, because in any case the net accretion rates of the galaxies present variations due to other effects and the particular formation and accretion history of each system.

\subsection{Inflow and outflow gas rates}

In this section, we calculate the average inflow and outflow rates previously discussed for the simulations including tracer particles.
In particular, we focus on G1, as 7 out of the 9 galaxies simulated with tracer particles belong to this group: Au6, Au9, Au13, Au16, Au23, Au24 and Au26.
The other two simulations -- Au5 and Au28 -- were excluded from the sample as they present strong perturbations during their evolution and no galaxies of G2 are present among the re-simulated ones.

\begin{figure*}
    \centering
    \hspace*{-0.4cm}
    \includegraphics[scale=1.15]{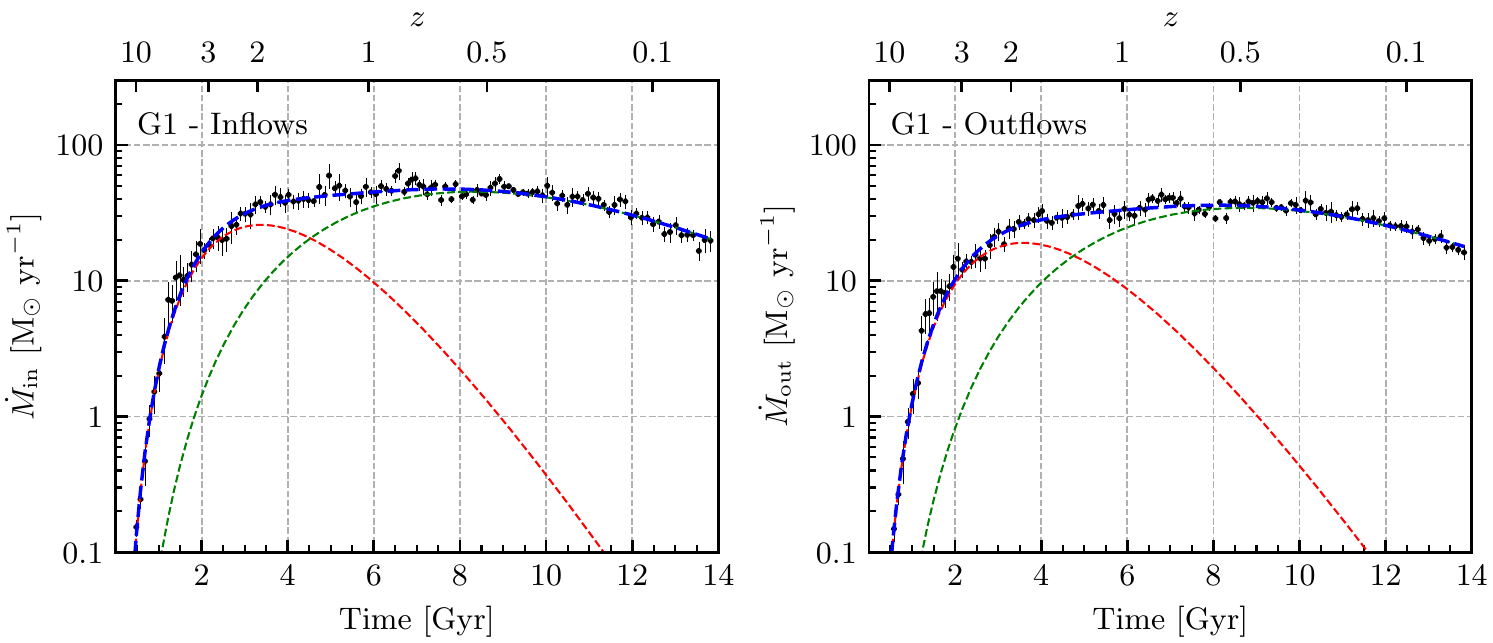}
    \caption{Average inflow (left panel) and outflow rate (right panel) for the galaxies of group G1, comprised of ``well-behaved'' galaxies. Error bars indicate plus or minus one standard deviation of the mean. The blue line in each panel indicates a fit using a double Schechter function, while the green and red lines shows the two functions used in said fit.}
    \label{fig:accretion_tracers_groups}
\end{figure*}

The average inflow and outflow rates of G1 display a similar behaviour, as can be seen from Fig.~\ref{fig:accretion_tracers_groups}, showing the usual trend of inflows being systematically at higher values than outflows.
Compared to the net accretion rates of G1 (Fig.~\ref{fig:net_accretion_cells_groups}), the early increase detected for the inflow and outflow rates is faster: at times $<2~\mathrm{Gyr}$, inflow accretion levels are $\sim 15~\mathrm{M}_\odot \, \mathrm{yr}^{-1}$ while outflow levels are $\sim 10~\mathrm{M}_\odot \, \mathrm{yr}^{-1}$.
We also find that the inflow and outflow rates remain approximately constant between $\sim 4~\mathrm{Gyr}$ and $\sim 10~\mathrm{Gyr}$, and decay thereafter.
The maximum values reached are $\sim 50~\mathrm{M}_\odot \, \mathrm{yr}^{-1}$ (at $\sim 6.5~\mathrm{Gyr}$) for inflows and $\sim 40~\mathrm{M}_\odot \, \mathrm{yr}^{-1}$ (at $\sim 7.0~\mathrm{Gyr}$) for outflows.

In this case, it is not possible to use a simple Schechter function for the whole evolution or an exponential fit for late times, as in the case of the net accretion.
However, a good fit is obtained if we use a double Schechter function of the form
\begin{equation*}
    \dot{M} = A \left[ \left( \frac{t}{\tau_1} \right)^{\alpha_1} \exp \left( - \frac{t}{\tau_1} \right) + \left( \frac{t}{\tau_2} \right)^{\alpha_2} \exp \left( - \frac{t}{\tau_2} \right) \right],
\end{equation*}
where we kept a single amplitude and left the other parameters free for simplicity.
It is worth noting that, although 5 five parameters might seem excessive to fit the inflow/outflow rates, these can be interpreted as coming from two different regimes during the formation of the galaxies: a first period, characteristic of the formation of the bulge (as shown above, the galaxies are not yet disc-like at high redshift), and a second phase which follows the more stable phase related to the formation of the disc component.
The results obtained using a non-linear least squares method for the fit are shown in Table~\ref{table:fit_parameters_schechter_reruns}, and the resulting functions are shown in red and green dashed lines in Fig.~\ref{fig:accretion_tracers_groups}. The uncertainties presented in the table correspond to the one standard deviation errors, which, in turn, were obtained as the square root of the diagonal elements of the estimated covariance matrix.

\begin{table}
    \centering
    \caption{Parameters of the fit using a double Schechter function for the average inflow and outflow accretion rates of the rerun simulations that belong to G1. These fits are shown in blue in Fig.~\ref{fig:accretion_tracers_groups}.}
    \begin{tabular}{lcc}
    \hline
    Parameter & Inflows & Outflows \\
    \hline
    $A$ [$\mathrm{M}_\odot\,\mathrm{yr}^{-1}$] & $1.5 \pm 0.4$ & $1.0 \pm 0.3$ \\ 
    $\alpha_1$ & $4.8 \pm 0.2$ & $5.0 \pm 0.2$ \\
    $\tau_1$ [Gyr] & $0.7 \pm 0.1$ & $0.7 \pm 0.1$ \\
    $\alpha_2$ & $5.2 \pm 0.2$ & $5.3 \pm 0.2$ \\
    $\tau_2$ [Gyr] & $1.6 \pm 0.1$ & $1.6 \pm 0.1$ \\
    \hline
    \end{tabular}
    \label{table:fit_parameters_schechter_reruns}
\end{table}

The best-fit parameters obtained for the inflow and outflow rates are similar in all cases, indicating that their time-evolution is similar, despite their different levels, reflected in the higher value of $A$ obtained in the case of the inflow rates.
The time corresponding to the maximum rates is $7.8~\mathrm{Gyr}$ for the inflows and $7.7~\mathrm{Gyr}$ for the outflows, and the corresponding maximum rates are $45.6~\mathrm{M}_\odot\,\mathrm{yr}^{-1}$ and $36.3~\mathrm{M}_\odot\,\mathrm{yr}^{-1}$, respectively.
These results confirm our findings of the previous Section, where we showed the close connection between the inflow and outflow rates onto the disc region.

\section{Discussion and conclusions}
\label{sec:conclusions}

In this work we have investigated the inflow, outflow and net accretion rates onto the discs of spiral galaxies, using simulations of galaxy formation in a cosmological context.
We used 30 high-resolution, zoom-in simulations from the Auriga Project \citep{Grand2017}, run with the moving-mesh magnetohydrodynamical code \textsc{arepo} \citep{Springel2010}.
The simulated galaxies have, at $z=0$, a virial mass similar to the MW and are relatively isolated, with no other massive galaxy in their immediate surroundings.
We complemented the sample with nine reruns, comprising a subset of the galaxies originally simulated, which include a treatment for tracer particles, allowing to follow the trajectories of gas elements in time.

The majority of the simulated galaxies have a well-defined, disc-like component at $z=0$, and the disc can be identified from $\sim 2$--$4~\mathrm {Gyr}$ on.
The evolution of the discs is diverse, in terms of their formation time and the occurrence of partial/total destruction events due to mergers and interactions with satellite systems.
In order to properly estimate the inflow, outflow and net accretion rates onto the discs, we calculated the disc radii and heights for all galaxies and times, using a criterion based on the distribution of the stellar mass to identify the physical region characterised by rotational motion.
Once the discs of the simulated galaxies were identified, we calculated the net accretion rates onto the disc region, as well as the inflow and outflow rates in the case of the nine resimulations for which a tracer particle treatment was available.

We calculated the net accretion rates for the full Auriga sample as the difference in the gas mass present in the discs between consecutive snapshots divided by the corresponding time interval (and properly considering star formation activity).
This approximation was shown to be adequate and was validated by comparing the results with those from the simulations with tracer particles.
We found that, in the DHI, inflows are more frequent and have higher levels compared to ouflows: the latter are detected for $\lesssim 30\%$ of the data points, and the outflow levels are typically 25\% lower than the inflow rates.
These results indicate that, for most of the evolution, there is a net inflow of gas onto the disc region.
In general terms, the net accretion rates integrated onto the discs are similar for all galaxies, exhibiting a rapid increase at early times, up to $\sim 2~\mathrm{Gyr}$, followed by an intermediate period of still increasing rates between $\sim 2$ and $6~\mathrm{Gyr}$.
After this time, we find more variations, with net accretion rates showing both increasing and decreasing patterns.

Galaxies with decreasing net accretion rates at late times were grouped together and identified as MW analogues, as all of them have smoothly growing, stable discs for at least the last $8~\mathrm{Gyr}$, with no significant perturbations.
In contrast, all galaxies characterised by increasing net accretion rates at late times have experienced important merger/interaction events after $\sim 6~\mathrm{Gyr}$, which lead to the partial/total destruction of their discs, even though most of them end up with well-formed disc-like components at $z=0$.
An exponential fit to the average net accretion rates of the latter group has been possible (starting at $\gtrsim 6~\mathrm{Gyr}$), and yielded a typical (negative) time-scale of $10.5~\mathrm{Gyr}$ 

In the following, we summarize our main results for the MW analogues.
In this case, it was possible to calculate the inflow and outflow rates separately, as 7 out of the 9 simulations including tracer particles belong to this group.
This allowed us to investigate additionally the relation between the inflow, outflow and star formation rates in the discs.

\begin{itemize}
    \item The net accretion rates of MW analogues, integrated over the discs, increase up to $\sim 6~\mathrm{Gyr}$ reaching an average maximum value of the order of $10~\mathrm{M}_\odot \, \mathrm{yr}^{-1}$, followed by an exponential-like decay up to the present time. An exponential fit to the late evolution (i.e. for times larger than $6\,\mathrm{Gyr}$) was done for the average relation, yielding a typical time-scale of the decay of $7.2~\mathrm{Gyr}$ and an amplitude of $33.1~\mathrm{M}_\odot \, \mathrm{yr}^{-1}$. The average net accretion rate of this group can also be fitted, during the whole evolution, with a Schechter function.

    \item The average inflow and outflow rates onto the discs show a rapid increase at early times, stay approximately constant between $\sim 4$ and $8~\mathrm {Gyr}$, and decay thereafter up to the present time. An exponential is not a good fit for the late evolution, but the whole evolution can be well fitted by a double Schechter function, a first one dominant at early times -- which can be interpreted as corresponding to the bulge formation period -- and a second one characterizing the late evolution -- i.e. the disc formation/evolution times.

    \item The ratio between the outflow and inflow rates is similar for all galaxies, and stays approximately constant over time, with a value of the order of 0.75. This indicates that the gas mass involved in outflows is approximately 25\% lower compared to the mass related to inflows.

    \item The ratio between the SFR and the inflow rate varies between 0.1 and 0.3 for all galaxies, with a general tendency to increase with time, indicating that, at all times, about 10--$30\%$ of the gas mass in the discs is converted into stars.

    \item The outflow rate over SFR ratio, the so-called mass-loading factor $\eta$, is found to be approximately constant during the evolution, with similar values for all galaxies. The median $\eta$ is of the order of 3.5--5.5, indicating that the feedback produced by the formation of stars is able to heat and pressurize the gas producing mass-loaded winds in the disc region. The $\eta$ values obtained, similar to those obtained with other simulation codes, are larger than observational estimates. This might indicate that simulations predict too strong winds, although it is also possible that observations do not properly trace all outflowing mass (see \cite{Kelly2022} for a comparison between the different feedback implementations).
\end{itemize}

The correlations found between the inflow/outflow rates and the SFRs in the discs reflect the inter-relation between the process of gas accretion -- which contributes fresh gas to the disc to form new stars -- and the generation of outflows which follow star formation.
Although these processes interact in a non-trivial manner and are affected by other processes occurring in the central regions of galaxies, such as black hole feedback, the tight correlations found for all galaxies and all times indicate that gas accretion is a key factor in the determination of the SFR in the discs, having a significant role in the circulation of gas in the DHI.

It is worth noting that quantifying the accretion rates onto the disc region of spiral galaxies using cosmological simulations is of utmost importance, not only to understand the process of circulation of gas in relation to inflows, star formation and outflows, but also to provide realistic accretion laws that can be used in more simplified models of galaxy formation.
For example, CEMs aiming at describing the properties of our Galaxy need to assume an accretion law, with corresponding time- and radial-dependencies.
In this paper, we showed that the net accretion rate for a relatively large sample of stellar discs belonging to galaxy haloes with masses 1--$2 \times 10^{12} \, \mathrm{M}_{\odot}$ simulated within a cosmological scenario is similar to that of the MW (i.e. for times characteristic of the formation of the discs) and can be well approximated by an exponential law.
In a second paper of this series, we will investigate the radial dependencies of the accretion rates, in order to investigate whether the inside-out behaviour typically assumed in CEMs can be taken as a robust prediction of cosmological simulations.

\section*{Acknowledgements}

CS and SEN acknowledge funding from Agencia Nacional de Promoci\'on Cient\'{\i}fica y Tecnol\'ogica (PICT-201-0667).
FAG acknowledges financial support from FONDECYT Regular 1211370, and from the Max Planck Society through a Partner Group grant.
FAG gratefully acknowledges support by the ANID BASAL project FB210003.

\section*{Data availability}

The scripts and plots for this article will be shared on reasonable request to the corresponding author.
The \textsc{arepo} code is publicly available \citep{Weinberger2020}.



\bibliographystyle{mnras}
\bibliography{bibliography} 




\appendix

\section{Dependence of radius, height, and accretion rates on enclosed mass fraction}
\label{sec:dependence_with_enclosed_mass}

\begin{figure}
    \centering
    \begin{tabular}{c}
        \includegraphics[scale=.9]{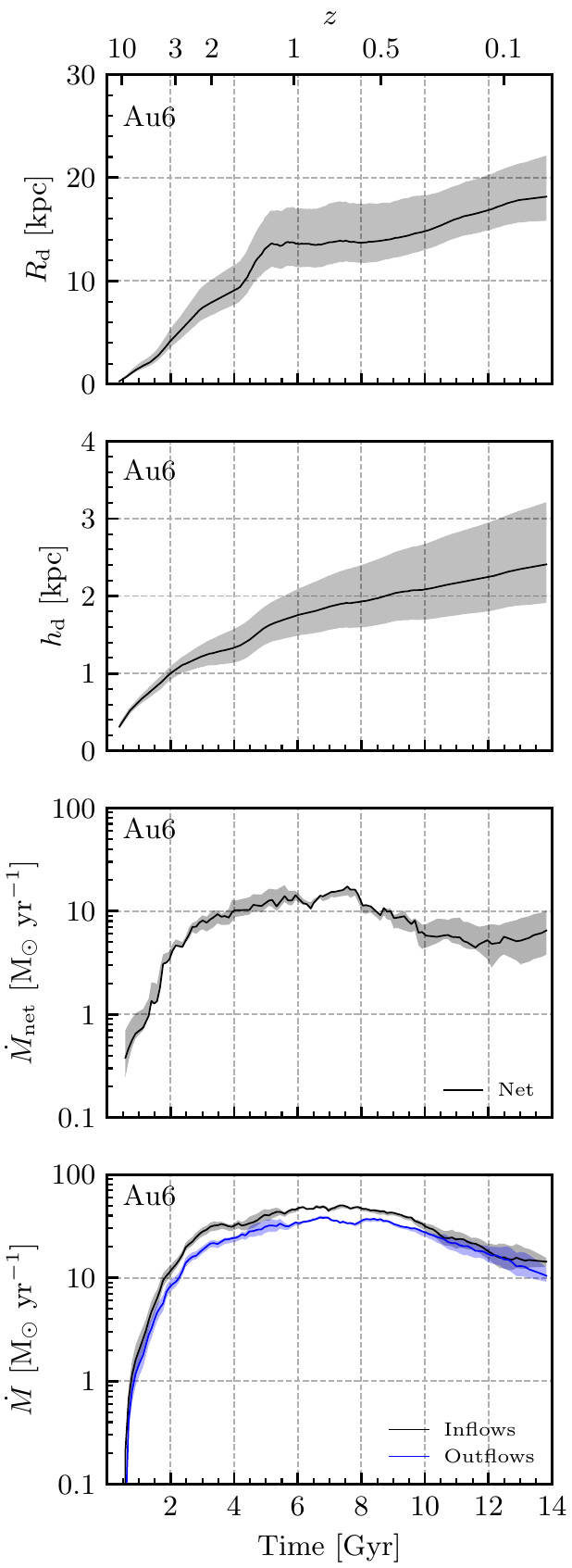}
    \end{tabular}
    \caption{
    Variation of the quantities analysed in this work based on the fraction of stellar mass enclosed in the disc region.
    We show changes for the disc radius (first panel from top to bottom), the disc height (second panel), the inflow-dominated regime for the net accretion rate (third panel) and the inflow and outflow rates (bottom panel) for galaxy Au6.
    Each panel shows a black line that corresponds to the standard mass fraction of 0.9. The grey regions plotted behind the curve show the range of possible values obtained if the mass fraction varies in the range $0.85-0.95$.
    Note that, although the disc radius and height increases with mass fraction, that is not necessarily the case for the accretion rates.
    }
    \label{fig:variations}
\end{figure}

As described in Section~\ref{sec:disc_radius_height}, we adopted simple criteria to define the disc radii and heights for all galaxies and times, allowing us to identify the region characterised by rotational motion.
In particular, we chose the disc radius and height to be that enclosing 90\% of the stellar mass ($f_\mathrm{m}=0.9$) for both the radial and vertical distributions, respectively.
We tested the mass fraction dependence of the disc radii and heights, as well as the net accretion rates, and found that the chosen $f_\mathrm {m}$ value does not significantly affect our results, as long as it allows the inclusion of a high fraction of the stellar mass.
Fig.~\ref{fig:variations} shows, for galaxy Au6, the variations in the disc radius, the disc height, the net accretion rate as a function of time and the evolution of the inflow and outflow rates, for our fiducidal choice of $f_\mathrm{m}=0.9$ (black lines), as well as the variations when we alternatively use $f_\mathrm{m}=0.85$ and $f_\mathrm{m}=0.95$ (grey shadows).
In the case of the disc radius, when changing $f_\mathrm{m}=0.9$ to 0.95 (0.85), we find a median increase (decrease) of 26\% (15\%) in $R_{\mathrm d}$.
Given that the gas density decreases with the distance to the center of the galaxy, it is expected to observe higher variations in the results when higher $f_\mathrm{m}$ values are assumed.
A similar result is found for the disc height, with a median increase (decrease) of 21\% (16\%) when varying $f_\mathrm{m}$ from 0.9 to 0.95 (0.85).
For both the disc height and radius, there is no observed change in the structure of the time-evolution.

The lower panels indicate the corresponding variations of the net accretion, and inflow and outflow rates when we vary $f_\mathrm{m}$.
The median change in the accretion rates can vary between $-14\%$ to 21\% for the net accretion, between $-10\%$ to 10\% for the inflows, and between $-9\%$ to 9\% for the outflows.

We therefore conclude that the results obtained throughout this paper are robust and are not significantly affected by our definition or choice of $f_\mathrm{m}$.

\section{Validating the results of the net accretion rates using cells}
\label{app:net_comparison}

As discussed above, the calculation of the net accretion rates applied to the full Auriga sample (Eq.~\ref{eq:net_acc_cells}) is an approximation, as particle trajectories can not be followed in these simulations.
The runs with tracer particles, on the other hand, allow a proper calculation of the inflow and outflow rates, and the net accretion rate is simply the difference between the two (Eq.~\ref{eq:net_acc_tracers}).

\begin{figure*}
    \centering
    \includegraphics{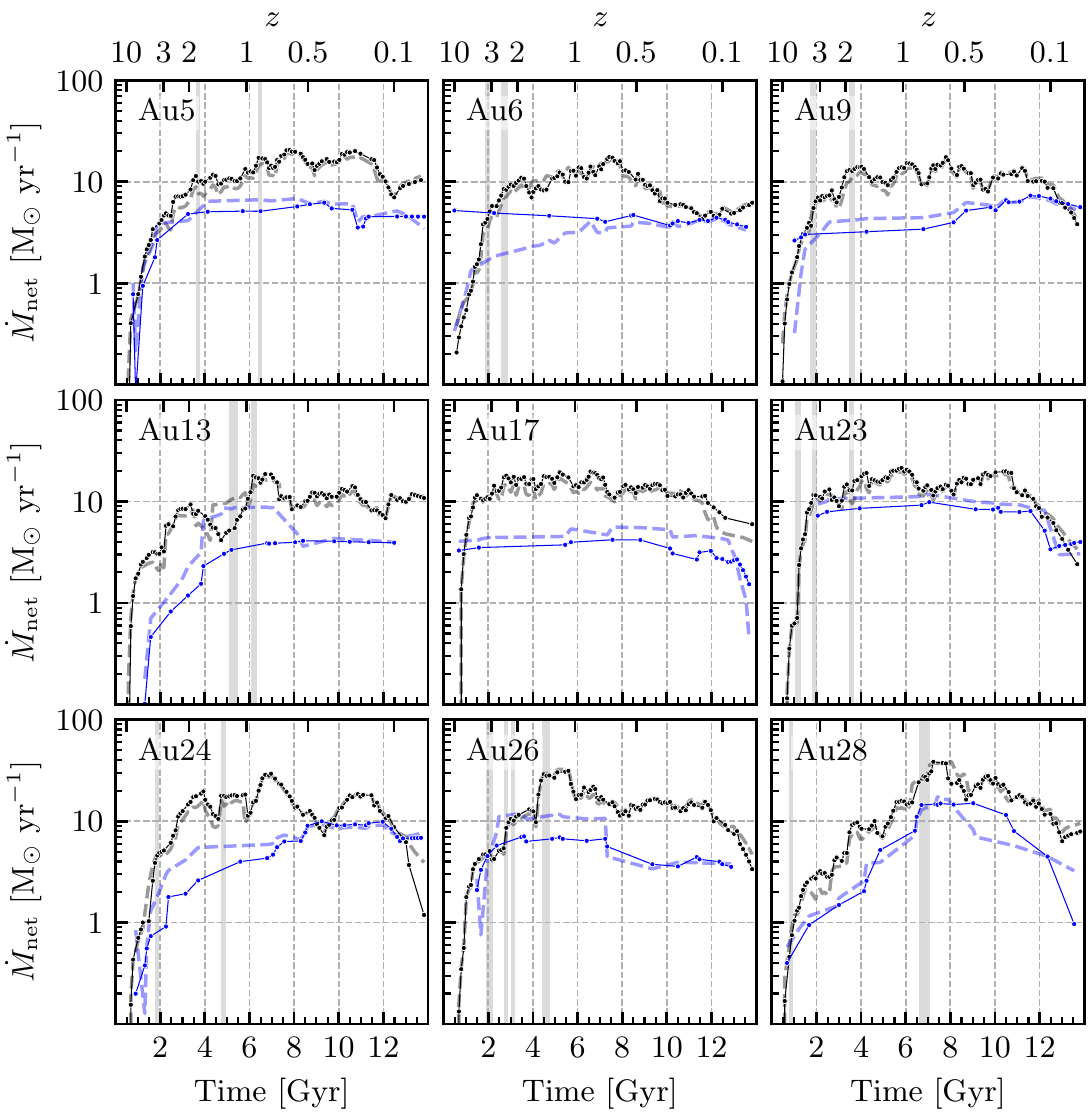}
    \caption{
    Comparison between the net accretion rates obtained using cells (dashed lines) and tracer particles (solid lines with circles) for the inflow-dominated (black) and outflow-dominated (blue) regimes.
    Shaded regions show  times where satellites with  $f_\mathrm{sat} = \frac{M_\mathrm{sat}}{M_\mathrm{cen}} > 0.1$ are present within $R_{200}$.
    The good agreement observed between the two methods indicates that it is possible to estimate the net accretion rate using the information provided by the gas cells.
    }
    \label{fig:net_accretion_tracers_reruns}
\end{figure*}

In order to validate the results found using the cell information, which allows the calculation of $\dot{M}_\mathrm{net}$ for the full Auriga sample, we compare the net accretion rates obtained with the two methods, for the 9 simulations for which we have the standard runs and those with tracer particles.
Fig.~\ref{fig:net_accretion_tracers_reruns} shows such a comparison for the case of inflow-dominated times (black lines) and outflow-dominated times (blue lines).
Solid lines correspond to our results calculated using tracers while dashed lines represent those for the calculations using the cell information.

We find a very good agreement for all galaxies and during most times between the two calculation methods, particularly in the case of the inflow-dominated times which comprise the majority of the data points.
The most important differences are detected in the case of Au13 between 4 and $6~\mathrm{Gyr}$.
This occurs right after a strong perturbation experienced by this galaxy, resulting from the approach of a satellite system (as indicated by the grey shades).

In the case of the outflow-dominated times, we detect more variations; however, it is important to note that the sampling is poor, as less than 30\% of the data points are in this regime.
In any case, for most of the galaxies, the differences are not significant except for limited periods of time.
The largest differences are detected for Au6 at times $\lesssim 6~\mathrm{Gyr}$, and for Au13 and Au26 at intermediate times.

These results show that it is possible to reliably estimate the \emph{net accretion rates} onto the discs using only the information provided by the gas cells in the Auriga simulations, particularly in the case of the inflow-dominated times, allowing a reliable estimation of the net accretion rates of the full galaxy sample.

\bsp	
\label{lastpage}
\end{document}